\documentclass{article} 

\usepackage{graphicx}		
\usepackage{latexsym}
\usepackage{amsmath, amssymb} 
\usepackage{setspace}
\usepackage[margin=0.8in]{geometry}
\usepackage{lineno}
\usepackage{natbib}
\setcitestyle{aysep={},round}

\begin{document}


\title{A point process framework for modeling electrical stimulation of the auditory nerve}

\author{Joshua H. Goldwyn$^{1,2,3}$    \and
        Jay T. Rubinstein$^{4,5,6}$ \and
        Eric Shea-Brown$^{1,7}$ }
\date{\today}	

\maketitle

\noindent
$^1$ Department of Applied Mathematics, University of Washington  \\
$^2$ Center for Neural Science, New York University \\
$^3$  Courant Institute of Mathematical Sciences, New York University \\
$^4$ Department of Biomedical Engineering, University of Washington \\
$^5$ Department of Otolaryngology, University of Washington \\
$^6$ Virginia Merrill Bloedel Hearing Research Center, University of Washington\\
$^7$ Program in Neurobiology and Behavior, University of Washington

\begin{abstract}
Model-based studies of auditory nerve responses to electrical stimulation can provide insight into the functioning of cochlear implants.  Ideally, these studies can identify limitations in sound processing strategies and lead to improved methods for providing sound information to cochlear implant users.   To accomplish this, models must accurately describe auditory nerve spiking while avoiding excessive complexity that would preclude large-scale simulations of populations of auditory nerve fibers and obscure insight into the 
mechanisms that influence neural encoding of sound information.  In this spirit, we develop a point process model of the auditory nerve that provides a compact and accurate description of neural responses to electric stimulation.  Inspired by the framework of generalized linear models, the proposed model consists of a cascade of linear and nonlinear stages.  We show how each of these stages can be associated with biophysical mechanisms and related to models of neuronal dynamics.  Moreover, we derive a semi-analytical procedure that uniquely determines each parameter in the model on the basis of fundamental statistics from recordings of single fiber responses to electric stimulation, including threshold, relative spread, jitter, and chronaxie.  The model also accounts for refractory and summation effects that influence the responses of auditory nerve fibers to high pulse rate stimulation.  Throughout, we compare model predictions to published physiological data and explain differences in auditory nerve responses to high and low pulse rate stimulation.  We close by performing an ideal observer analysis of simulated spike trains in response to sinusoidally amplitude modulated stimuli and find that carrier pulse rate does not affect modulation detection thresholds.  

\end{abstract}

\thispagestyle{empty}

\newpage
\twocolumn
\setcounter{page}{1}

 
\section{Introduction}
\label{sec:introduction}

Cochlear implants restore a sense of hearing to individuals with severe to profound hearing loss by electrically stimulating the primary sensory neurons in the auditory pathway.  A complete understanding of the transformation from electrical stimulation to neural responses would aid the design of improved cochlear implant sound processing strategies.  Computational and mathematical models, especially those that are carefully constrained by available neurophysiological data, can play an essential role in exploring this transformation. In addition to providing an efficient platform for testing and evaluating stimulation paradigms, models provide quantitative tools for studying how neural mechanisms influence the transmission of sound information in the auditory system.

In this paper, we develop a point process modeling framework that can be used to simulate the response of the auditory nerve (AN) to cochlear implant stimulation.  The dynamical and stochastic features of the model are matched to statistics that characterize neural responses to single and paired pulses of electrical current --  stimuli that are commonly used to characterize the response properties of AN fibers in animal models of electric hearing \citep{Hartmann1984, Vandenhonert1984, DynesThesis, Bruce1999single, Miller1999, Shepherd1999,  Cartee20001, Miller2001, Miller2001bi, Cartee2006}.   We go on to show that this model can provide insight into the responses of neurons to extended pulse trains of electrical stimulation.  Point process models also have mathematical properties that facilitate analyses of auditory signal detection \citep{Heinz2001a, Heinz2001b}, and we close by relating results from our model to the psychophysical test of amplitude modulation detection.

Our approach connects two prior modeling frameworks.  The first is the stochastic threshold crossing model \citep{Bruce1999train, Bruce1999single}.  This model is a useful phenomenological representation of AN spiking. A number of cochlear implant psychophysics experiments have been studied with this model \citep{Bruce1999intensity, Xu2007, Goldwyn2010jcn}.  Unfortunately, as stated by its creators, it lacks sufficient dynamical details to provide valid predictions of AN responses to high rate stimulation \citep{Bruce1999train}.  Studying responses to high pulse rate stimulation is necessary to characterize contemporary cochlear implant stimulation strategies, and for reasons that we discuss in more detail below, is a primary focus of this work.  We therefore extend the stochastic threshold crossing model by turning to a more general class of point process models.  

Point processes describe spiking via an instantaneous firing rate that varies over time \citep{Perkel1967, Johnson1996, Truccolo2005}. They have been frequently applied to model auditory nerve firing  \citep{Miller1992, Litvak2003sinusoid, Trevino2010, Plourde2011}.   Our approach is largely motivated by a specific family of point processes: generalized linear models (GLMs) \citep{Paninski2004pp, Truccolo2005}.  GLMs have emerged as an essential tool for modeling physiological data and investigating the coding and computational properties of neurons \citep{Paninski2004pp, Paninski2007}, including auditory neurons \citep{Trevino2010, Plourde2011}.  Moreover, they hold particular promise for applications to sensory neural prostheses because they have useful mathematical properties that permit efficient parameter fitting to spike train data \citep{Paninski2004pp}, they can be used to optimally decode spike trains \citep{Paninski2007}, and they can be used in connection with real-time optimization methods to identify stimulation patters that control the timing of evoked spikes \citep{Ahmadi2011}.  

We apply our point process model to investigate differences between AN responses to low and high pulse rate electric stimulation.  Interest in high pulse rate stimuli arises from the intuitive notion that higher stimulation rates should provide greater temporal information to cochlear implant listeners and thereby improve speech reception.  Indeed, computational modeling and neurophysiological studies have indicated that stimulating neurons with 5000 pulses per second (pps) pulse trains can desynchronize neural responses \citep{Rubinstein1999, Litvak2001, Litvak2003desynch}. This may return AN activity to a state that more closely resembles normal spontaneous activity in the healthy cochlea \citep{Rubinstein1999, Litvak2003desynch}.  In related modeling and experimental studies, high pulse rate stimulation has been shown to produce neural firing rates that more faithfully reproduce the envelope-s of temporally-modulated stimuli \citep{Litvak2003vowel, Litvak2003sinusoid, Chen2007, Mino2007}.  Psychophysical data from tests of amplitude modulation detection, however, have not yielded clear indications that high pulse rate stimulation improves the ability of listeners to detect temporal modulation \citep{Galvin2005, Pfingst2007, Galvin2009}.  More generally, speech tests have reported mixed results as to whether higher pulse rate stimulation improves speech reception in cochlear implant listeners \citep{Brill1997, Kiefer2000, Loizou2000, Vandali2000, Holden2002, Friesen2005}. 
Computational modeling can help trace the connections between the perceptual benefits, if any, of high pulse rate stimulation and the neural responses that it evokes.

\section{Method}
\label{sec:method}

In this section we introduce the point process framework and discuss how it can be parameterized using commonly reported statistics of responses to single pulse and paired pulse stimuli.  These include threshold, relative spread, jitter, chronaxie, summation threshold, and two refractory effects.  We begin by introducing the response statistics that we used to construct the model (Sec.~\ref{subsec:responsestats}), followed by a general discussion of how these statistics can be extracted from a point process model (Sec.~\ref{subsec:pptheory}).  We then introduce the specific model framework that we will use throughout the study (Sec.~\ref{subsec:formulation}) and explain the procedure by which each parameter in the model can be uniquely associated with a response statistic (Sec.~\ref{subsec:parameterization}).

\subsection{Response statistics}
\label{subsec:responsestats}

A first-order description of neuronal excitability in response to electrical stimulation is the \emph{firing efficiency curve}.  This is an input/output function that relates the current level of a single pulse of current to the probability that the stimulus evokes a spike.  As shown by Verveen in his pioneering recordings of green frog axons \citep{Verveen1960}, the firing efficiency curve can take the form of a sigmoid function and be approximated by the integral of a Gaussian distribution.  The median of the underlying Gaussian distribution represents the stimulus level that elicits a spike with probability one-half.  This level is referred to as the \emph{threshold} of the neuron, which we denote by $\theta$.  A measure of the variability of spike initiation is the \emph{relative spread (RS)}, defined as the standard deviation of the underlying Gaussian distribution divided by its mean.  Firing efficiency curves are frequently used to characterize the response of AN fibers to cochlear implant stimulation in electrophysiological experiments \citep{DynesThesis, Bruce1999single, Miller1999, Shepherd1999, Miller2001bi} and to parameterize computational models \citep{Bruce1999single, Stocks2002, Macherey2007, Imennov2009, OGorman2009}.  We will follow this approach and use the firing efficiency curve, in particular the parameters $\theta$ and RS, as the starting point for constructing our point process model.

The firing efficiency curve summarizes the excitability of a neuron in response to a single pulse of stimulation.  For longer pulse durations, the threshold current level is typically smaller, due to the capacity of the neural membrane to integrate charge over time.  This dependence of threshold on pulse duration can be summarized by the \emph{chronaxie};  the pulse duration at which the threshold level is twice what it would be for a much longer pulse duration (how long, and therefore reported values of chronaxie, vary in different experiments). This basic feature of membrane dynamics is absent from earlier stochastic threshold crossing models \citep{Bruce1999train, Litvak2003sinusoid}.

The firing efficiency curve encapsulates variability in the initiation of spikes (or lack thereof), but an additional source of randomness is in the timing of spikes.   The standard deviation of spike times evoked by a single pulse is known as \emph{jitter}. It is another statistic that has been widely studied in physiological and modeling studies of electrical stimulation of the auditory nerve \citep{Vandenhonert1984, Miller1999, Javel2000, Miller2001bi, Cartee2006}.  Jitter can depend on pulse duration and pulse level, but in our model we will only use the value of jitter that is most commonly reported in the literature: the value measured for the same pulse duration and level that defines the threshold.

In order for the model to generate realistic responses to high pulse rate stimulation, it must also include spike history effects.  The first, and most essential, is a refractory effect that reduces the excitability of the model neuron in the immediate aftermath of an evoked spike.  This effect will be implemented as a transient increase in the threshold value  $\theta$ following a spike, similar to the method of Bruce and colleagues \citep{Bruce1999train}. The point process will also include a second spike history effect -- a transient increase in RS immediately after a spike \citep{Miller2001}.  Based on simulation results from a biophysically detailed computational model of an AN fiber, it has been hypothesized that this phenomenon is a signature of the random activity of ion channels in a cell membrane known as channel noise \citep{Matsuoka2001, Mino2006}.  Modeling channel noise in the auditory nerve in the context of cochlear implant stimulation is important for a number of reasons.  First, neurons in the deafened cochlea do not receive the typical synaptic input from inner hair cells, and therefore ion channels may generate a dominant noise source in this stage of auditory processing.  Second, channel noise may be a mechanism by which high pulse rate stimulation can desynchronize neural responses, thereby leading to improved encoding of electrical stimuli \citep{Rubinstein1999, Litvak2003vowel, Litvak2003sinusoid, Mino2007}.   

Finally, we incorporate an additional feature into the model that is relevant for high carrier pulse rates -- the capacity of a neuron to integrate consecutive subthreshold pulses.  In other words, the model will account for the phenomenon by which multiple pulses are more likely to evoke a spike than would be expected if each pulse acted independently on the neuron.  This effect, known as \emph{summation} or \emph{facilitation}, has been observed in responses to pairs of pulses with short interpulse intervals \citep{Cartee20001, Cartee2006} and to high rate pulse train stimuli \citep{Heffer2010}.  As we will see, this form of interpulse interaction creates fundamental differences between responses to low and high rate pulse train stimulation.

\subsection{Point process theory}
\label{subsec:pptheory}

In this section, we introduce a number of basic ideas from the theory of point processes and illustrate their connections to the response statistics discussed above.  A point process model is completely defined by its \emph{conditional intensity function} \citep{Daley2003}, which can be interpreted as the instantaneous probability that a neuron will spike \citep{Truccolo2005}.  We denote the conditional intensity function as $\lambda(t|I,H)$, where $I$ is the stimulus applied to the neuron and $H$ represents the past history of the neuron, both of which should be viewed as functions of time.
A related and important quantity is the \emph{integrated intensity function}:
\begin{align}
\Lambda(t_1,t_2|I, H) = \int_{t_1}^{t_2} \lambda(s|I, H) ds. \notag
\end{align}
From  $\Lambda(t_1,t_2|I,H) $, one can define the probability that there will be a spike in a time interval $\left[t_1, t_2, \right]$.  This function is known as the \emph{lifetime distribution function} and is given by \citep{Daley2003}:
\begin{align}
\label{eq:Lifetime}
L(t_1,t_2|I, H) = 1 - e^{-\Lambda(t_1,t_2|I, H)}.
\end{align}

The lifetime distribution function is central to our analysis because it forms the mathematical link between the point process model and the statistics that describe the responses of neurons to single and paired pulses.  Consider, for instance, responses to a single pulse of current level $I$ and some duration $D$.   If we let $t_2$ in Eq.~\ref{eq:Lifetime} become sufficiently large, then the lifetime distribution function defines the probability with which a single pulse of a certain current level $I$ will ever evoke a spike.  In other words, when viewed as function of $I$, it is equivalent to the firing efficiency curve.  If we now change our perspective, and view Eq.~\ref{eq:Lifetime} as a function of $t_2$, for a fixed value of $I$, then this function gives the probability that a spike has been evoked before time $t_2$.  In this context, the lifetime distribution function describes the temporal dispersion of spike times, and can thus be used to calculate jitter.  We now present how the single pulse and paired pulse response statistics discussed in Sec.~\ref{subsec:responsestats} can formally be obtained using point process theory.

We begin with a set of measures that are obtained from responses to single pulses, and we therefore assume that they are not influenced by spike history effects.  This allows us to omit the term $H$ for now and let $I$ refer the current level of a single stimulating pulse. As introduced above, the threshold value $\theta$ is the current level at which the firing efficiency curve has a value of $1/2$.  From the definition of the lifetime distribution, $\theta$ must satisfy
\begin{align}
\frac{1}{2} = L(0,\infty | \theta). \notag
\end{align}
Applying Eq.~\ref{eq:Lifetime}, this can be rewritten in terms of the integrated intensity function:
\begin{align}
\label{eq:log2}
\Lambda(0, \infty | \theta) =  \log 2
\end{align}

To define RS, we generalize the traditional definition that is based on assuming the firing efficiency curve is shaped like the integral of a Gaussian distribution \citep{Verveen1960}.  From Eq.~\ref{eq:Lifetime}, it is apparent that the firing efficiency curve for a point process model is not necessarily an integrated Gaussian function.  It is, however, the cumulative distribution of some probability density function.  To be concrete, we define an associated probability density function as the derivative of the lifetime distribution function with respect to the current level $I$:
\begin{align}
l_I(0,\infty | I) = \frac{d}{d I} \left[   \Lambda(0, \infty| I ) \right] e^{-\Lambda(0, \infty| I)} \notag
\end{align}
By analogy with the traditional definition, we let RS be the standard deviation of this associated density function normalized by its mean.

To define chronaxie in terms of a point process model, we use Eq.~\ref{eq:log2} and compare threshold current levels for varying pulse durations.  Let $\theta(D_\infty)$ denote the threshold for a monophasic pulse of a long pulse of duration $D_\infty$. Then, the chronaxie is the phase duration $D_{c}$ for which the single pulse threshold $\theta(D_c)$ is twice the value of $\theta(D_\infty)$.  Specifically, $D_c$ must satisfy the following two conditions:
\begin{align}
&\theta(D_c) = 2\theta(D_\infty) \notag \\
&\Lambda(0,\infty | \theta(D_c)) = \Lambda(0,\infty | \theta(D_\infty)) = \log(2) \notag
\end{align}
 
The final single pulse statistic we consider is jitter.  As mentioned above, jitter can be obtained from the lifetime distribution function when it is viewed a function of time.  Following the standard practice, we will consider jitter in spike times evoked by a single pulse presented at the threshold level $\theta$.  Then the probability of at least one spike occurring in the interval $[0,t]$, conditioned on the event that the stimulus produces a spike at any time, is twice the lifetime distribution function: $2 L(0,t | \theta)$.  We define the associated probability density function by taking a derivative with respect to time of Eq.~\ref{eq:Lifetime}, and see that jitter is the standard deviation of the following density function for the probability that a spike occurs at time $t$:
\begin{align}
\label{eq:jitterpdf}
l_t(0,t | \theta) =  2  \lambda(0, t | \theta) e^{-\Lambda(0, t | \theta)}.
\end{align}

We now turn to responses evoked by pairs of pulses. Using a summation pulse paradigm \citep{Cartee20001, Cartee2006}, we can quantify how threshold current levels change for a stimulus consisting of two pulses of equal current levels, separated by an interpulse interval (IPI) of varying duration.  Note that in experimental studies, the question asked  is whether the pulse pair evokes a spike; the timing of the spike is not relevant.  Assuming that the response to the pulse pair is not influenced by any prior spiking activity, the summation threshold satisfies the same threshold condition given by Eq.~\ref{eq:log2}, the only changes are that the stimulus $I$ is a pair of pulses and $\theta$ is the current level at which this pair of pulses evokes at least one spike on half of all trials, on average.

Lastly, we observe how refractory effects can be incorporated by allowing the single pulse threshold $\theta$ and RS to depend on the time since the last pulse.  In the refractory pulse paradigm \citep{Cartee20001, Miller2001, Cartee2006}, a strong first pulse forces the neuron to spike and then a firing efficiency curve is measured from the responses to a second pulse presented some time later.  The mathematical relationships between that firing efficiency curve and the lifetime distribution function remain unchanged, but they must be formally modified by adding the spike history term $H$.

\subsection{Model formulation}
\label{subsec:formulation}

\begin{figure*}
\includegraphics[width=174mm]{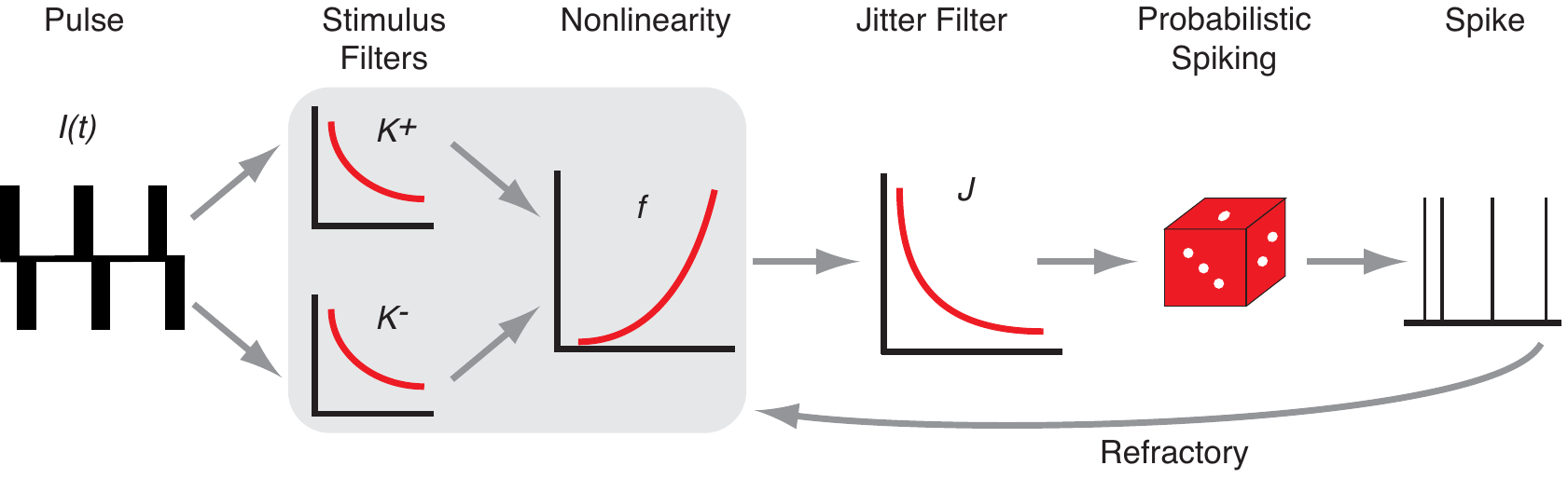} 
\caption{ Schematic diagram of the point process model.  Current input to the model, shown here as a train of biphasic pulses, is passed through a cascade of linear filters and a nonlinear function.  This produces a value -- the conditional intensity -- that defines the instantaneous probability of a spike, which is then used to generate random sequences of spike times.  Previous spikes provide feedback that modulates the stimulus filters and the nonlinearity. }
\label{fig:ModelCartoon}
\end{figure*}

We have presented the connection between the point process model and response statistics in a general manner.  In this section, we make these connections more explicit by proposing a specific structure for the conditional intensity function.  The model consists of a cascade of linear and nonlinear stages followed by a probabilistic spike generator.  This structure is inspired by the popular generalized linear model (GLM) class of point process neuron models.  Fig.~\ref{fig:ModelCartoon} illustrates the model features.  The model differs from standard GLMs in several ways.  These include a nonlinear stage that depends on the time since the last spike, an asymmetry in how the positive and negative phases of charge-balanced pulses are filtered, and a secondary filter that adds variability to simulated spike times. 

The action of the model, depicted schematically in Fig.~\ref{fig:ModelCartoon}, can be summarized as follows: an incoming pulse train $I(t)$ is passed through stimulus filters $K^+(t)$ and  $K^-(t)$, the outputs of the filters are recombined and used as input to a nonlinear function $f(\cdot)$.  To incorporate refractory effects, the stimulus filters and the nonlinearity are all modified depending on the time since the last spike.  An additional filter $J(t)$ is included to add variability to simulated spike times.  The result of this chain of events is an instantaneous, history-dependent value for the conditional intensity function that defines the point process model:
\begin{align}
\label{eq:cifmodel}
\lambda(t|I,H) = [J * f( K^+  * I^+ + K^-  * I^-)](t),
\end{align}
where $I^+$ and $I^-$ represent the positive and negative portions of the input and $*$ represents the convolution operator. 

A variety of methods exist for fitting GLMs to neural data including reverse correlation to white noise stimuli \citep{Simoncelli2004} and maximum likelihood methods \citep{Paninski2004pp}.  We pursue a different route and take advantage of the special structure of the proposed model and the mathematical relationships in Sec.~\ref{subsec:pptheory} to develop a semi-analytical procedure that uniquely identifies parameters in the point process model with specific response statistics.  For reasons of mathematical tractability and biological relevance, which we articulate more fully below, we define the components of the model as follows:
\begin{subequations}
\label{eq:model}
\begin{align}
f(v) &=  \left\{  \begin{array}{l} v^\alpha \quad \mbox{if }v\ge0  \\ 0 \qquad \mbox{else}  \end{array} \right. \label{eq:model1}  \\
K^+(t) &=  \left\{  \begin{array}{l}  \frac{\kappa}{\tau_K} e^{-t/\tau_K}  \quad \mbox{if }t\ge0  \\ 0 \qquad \qquad \mbox{else}  \end{array} \right. \label{eq:model2} \\
K(^-t) &=  \left\{  \begin{array}{l}  \frac{\beta \kappa}{\tau_K} e^{-t/\tau_K}  \quad \mbox{if }t\ge0  \\ 0 \qquad \qquad \mbox{else}  \end{array} \right. \label{eq:model3} \\
J(t) &=  \left\{  \begin{array}{l}  \frac{1}{\tau_J} e^{-t/\tau_J}  \quad \mbox{if }t\ge0  \\ 0 \qquad \qquad \mbox{else}  \end{array} \right. \label{eq:model4}  .
\end{align}
\end{subequations}
This model has five parameters: $\alpha$, $\kappa$, $\tau_K$, $\beta$, and $\tau_J$.  We next show how they are uniquely determined by the five response statistics discussed in Sec.~\ref{subsec:responsestats}: relative spread, threshold, chronaxie, jitter, and summation pulse threshold.  We will also illustrate how the model can account for refractory effects by allowing the parameters $\alpha$ and $\kappa$ to depend on the time since the previous spike.

\subsection{Model parameterization}
\label{subsec:parameterization}

We begin by discussing the response measures that do not depend on spike history, and thus neglect $H$ for now.  To simplify our presentation, we introduce some additional notation.  Let $w(t)$ be the waveform of the stimulus; for instance, the waveform of a monophasic pulse is a rectangular step that reaches a maximum value of 1.  Alternatively, the waveform function for a biphasic pulse consists of two rectangular phases, one reaches a value of $+1$ and the other $-1$.  We next define the filtered waveform function $W(t)$, which represents the action of $K^+$ and $K^-$ on $w(t)$:
\begin{align}
W(t) = \int_0^t \left[K^+(s) w^+(t-s)  + K^-(s) w^-(t-s) \right] ds, \notag
\end{align}
where $w^+$ and $w^-$ are the positive and negative parts of the waveform function.

For the case that all pulses have identical current level $I$, for instance in single pulse or summation pulse experiments, the intensity function for the point process model has the compact form:
\begin{align}
\lambda(t|I) = (\kappa I) ^\alpha  [J *W^\alpha](t). \notag
\end{align}
The parameter $\kappa$ and the pulse current level $I$ both factor out due to our choice of a power law nonlinearity.  In addition, the jitter filter $J(t)$ can be ignored when considering the integrated intensity function over all time.  This follows from Fubini's Theorem \citep{Jones2001} and the fact that $\int_0^\infty J(t)dt =1$:
\begin{align}
  \int_0^\infty[J *W^\alpha](t)dt =   \int_0^\infty W(t)^\alpha dt. \notag
\end{align}
If we denote this integral as
\begin{align}
\label{eq:defineWcal}
\int_0^\infty W(t)^\alpha dt.\equiv \mathcal{W_\alpha},
\end{align}
then the lifetime distribution function for the time interval $[0, \infty)$ can be written as:
\begin{align}
\label{eq:ModelLifeTime}
L(0,\infty | I)  = 1 - \exp \left( -\kappa^\alpha I^\alpha \mathcal{W_\alpha} \right). 
\end{align}
As noted above, this lifetime distribution function is, in the language of neurophysiology, the firing efficiency curve or input/output function of the neuron in response to a single pulse of applied current.  Next, we use this equation to parameterize the model.

\subsubsection{Relative spread}
\label{subsubsec:rs}

The choice of a power law nonlinearity significantly simplifies the parameterization process. The lifetime distribution function in Eq.~\ref{eq:ModelLifeTime} is a Weibull cumulative distribution function with shape parameter $\alpha$ and scale parameter $ \left[ \kappa \sqrt[\alpha]{ \mathcal{W_\alpha} }\right]^{-1} $.   We define the relative spread as the associated standard deviation divided by the mean. Using known expressions for these values we find \citep{Rinne2009}:
\begin{align}
\label{eq:rsfit}
RS =  \sqrt{ \frac{\Gamma({1+\frac{2}{\alpha}})}{\Gamma^2({1+\frac{1}{\alpha}})}-  1},
\end{align}
where $\Gamma(\cdot)$ is the Gamma function.  Remarkably, the relationship between RS and $\alpha$ does not depend on any other model or stimulus parameters. 
There is therefore a one-to-one relationship between the response statistic RS and the model parameter $\alpha$.  To fit the model, we invert this relationship to find $\alpha$ as a function of RS.  In practice this must be done numerically.

\subsubsection{Threshold}
\label{subsubsec:threshold}

The threshold current level $\theta$ is obtained by solving Eq.~\ref{eq:log2}, which is equivalent to finding the median of the Weibull distribution:
\begin{align}
\label{eq:thetafit}
\theta  = \frac{1}{\kappa} \sqrt[\alpha]{\frac{\log 2}{ \mathcal{W_\alpha} }}.
\end{align}
We invert this relationship in order to have an expression for the model parameter $\kappa$ as a function of the response statistic $\theta$:
\begin{align}
\label{eq:kappafit}
\kappa = \frac{1}{\theta} \sqrt[\alpha]{\frac{\log 2}{ \mathcal{W_\alpha} }}.
\end{align}
Note that $\kappa$ depends on the model parameters $\alpha$ as well as $\beta$, and $\tau_\kappa$ (through the definition of $\mathcal{W_\alpha}$ in Eq.~\ref{eq:defineWcal}).  As we will see, the values of these parameters are independent of $\theta$, and can therefore be viewed as (known) constants in this equation.

\subsubsection{Chronaxie}
\label{subsubsec:chronaxie}

In order to fit the model parameter $\tau_\kappa$, observe that the waveform function $w(t)$ and thus the associated function $\mathcal{W}_\alpha$ both depend on the pulse duration.  We denote this dependence as $\mathcal{W}_\alpha(D)$.  By definition, the chronaxie value is the pulse duration $D_c$ for which the threshold for a monophasic pulse is twice the value of the threshold in response to a monophasic pulse of a long duration $D_\infty$ (the exact value of $D_\infty$ varies across experimental studies).  We can use Eq.~\ref{eq:thetafit} to define the threshold current level as a function of pulse duration, and obtain the relationship: 
\begin{align}
\frac{1}{\kappa} \sqrt[\alpha]{\frac{\log 2}{ \mathcal{W}_\alpha(D_c) }} = \frac{2}{\kappa} \sqrt[\alpha]{\frac{\log 2}{ \mathcal{W}_\alpha(D_\infty) }}. \notag
\end{align}
The factors of $\kappa$ cancel one another and the equation can be further simplified to:
\begin{align}
\label{eq:chronaxiefit}
\frac{\mathcal{W}_\alpha(D_\infty)}{\mathcal{W}_\alpha(D_c)} = 2^\alpha. 
 \end{align}
Conveniently, the solution to this equation depends only on $\mathcal{W}_\alpha$ and $\alpha$.  Now, $\alpha$ can be assumed to have a known value since it is uniquely determined by RS (see Eq.~\ref{eq:rsfit}).  Furthermore, $\mathcal{W}_\alpha$ does not depend on $\beta$ since chronaxie is defined with respect to monophasic pulses.  As a consequence, the only undetermined model parameter in this equation is $\tau_\kappa$, which enters in the definition of $\mathcal{W}_\alpha$.  There is no analytical way to identify $\tau_\kappa$ as a function of chronaxie, but we can numerically evaluate $\mathcal{W}_\alpha(D_\infty)$ and  $\mathcal{W}_\alpha(D_c)$ and then use numerical root finding methods to determine the value of $\tau_\kappa$ that satisfies the relationship in Eq.~\ref{eq:chronaxiefit}, for given values of chronaxie $D_c$ and reference pulse duration $D_\infty$.

\subsubsection{Summation threshold}
\label{subsec:summationthreshold}

To parameterize the temporal summation properties of the model, we replicate the paired pulse experiments of \cite{Cartee20001} using pseudo-monophasic pulses \citep{Cartee20001, Cartee20002}.  In these experiments, two charge-balanced pulses are presented at varying interpulse intervals ($t_{IPI}$).  The threshold current level for the combined response to both pulses, which we denote $\theta_2$, is compared to the threshold current level for a single pulse presented in isolation, which we denote $\theta_1$.  \cite{Cartee20001} measured thresholds at three interpulse intervals (100, 200 and 300 $\mu s$) and summarized the relationship between $\theta_1$ and $\theta_2$ using the function:
\begin{align}
\label{eq:sumratio1}
\frac{\theta_2}{\theta_1} = 1 - \frac{1}{2} \exp{\left( -t_{IPI}/\tau_{sum} \right)}.
\end{align}
The parameter $\tau_{sum}$ controls the length of time over which the neuron can effectively sum consecutive subthreshold pulses.

To translate this into the language of the point process model, we use Eq.~\ref{eq:log2} to obtain a relationship between the single and paired pulse thresholds.  Let $\mathcal{W}_{\alpha,1}$ denote the filtered waveform for a single pulse and $\mathcal{W}_{\alpha,2}(t_{IPI})$ be the filtered waveform for a pair of pulses separated by an interpulse interval of length $t_{IPI}$.  Then the ratio of the thresholds can be written as:
\begin{align}
\label{eq:sumratio2}
\frac{\theta_2}{\theta_1} = \sqrt[\alpha]{\frac{\mathcal{W}_{\alpha,1}}{\mathcal{W}_{\alpha,2}(t_{IPI})}}
\end{align}

The model parameters that enter into this equation are $\alpha$, $\tau_\kappa$, and $\beta$ (since the pulses in this paradigm have negative phases).  We treat $\alpha$ and $\tau_\kappa$ as constants in Eq.~\ref{eq:sumratio2} since they do not depend on $\beta$ and can be determined at prior steps in the fitting procedure.  We determine $\beta$ by minimizing the mean square error between the two representations of $\theta_2/\theta_1$ on the right hand sides of Eqs.~\ref{eq:sumratio1} and~\ref{eq:sumratio2}.
This must be done using numerical methods for integration and minimization. The parameter $\beta$ will take a value between zero and one, with smaller values leading to greater subthreshold integration of consecutive pulses.  We are are not aware of experimental studies that that indicate how the temporal integration properties of the AN change immediately after a spike, so $\beta$ is left as a constant that does not depend on spike activity.

\subsubsection{Jitter}
\label{subsubsec:jitter}

The remaining undetermined parameter is $\tau_J$.  To determine its value, we recall that jitter in the point process model is defined as the standard deviation of the probability density function in Eq.~\ref{eq:jitterpdf}.  Once again, we cannot obtain a simple analytical relationship between $\tau_J$ and jitter, but we can use numerical integration and root finding methods to find the value of $\tau_J$ for which the standard deviation of the density function in Eq.~\ref{eq:jitterpdf} is equal to a measured value of jitter.  In other words, we solve the minimization problem:
\begin{align}
\underset{\tau_J}{\operatorname{argmin}} \left| \sqrt{ \int_0^\infty t^2 l_t(0,t | \theta) dt - \left( \int_0^\infty t l_t(0,t | \theta) dt \right)^2} - \gamma \right|, \notag
\end{align}
where $\gamma$ is the experimentally observed jitter value.
Although the density function $l_t(0,t | \theta) $ depends on all of the other parameters in the model, by this step in our parameterization method all other parameter values will have been specified and we can therefore treat them as (known) constants when determining the value of $\tau_J$.   Available data from cat AN fibers did not provide strong evidence that jitter changes significantly due to spike history effects \citep{Miller2001}.  Thus, the parameter $\tau_J$ does not depend on spike activity and is held constant throughout our simulations.

\subsubsection{Refractory effects}
\label{subsubsec:Refractory}

Finally, we describe how spike history effects can be incorporated in the model by allowing the parameters $\kappa$ and $\alpha$ to depend on the time since the last spike.  
The relationship between $\kappa$ and $\theta$ in Eq.~\ref{eq:kappafit} shows that increasing $\kappa$ immediately following a spike allows the model to exhibit a common refractory property, whereby neurons are less excitable in the aftermath of a spike.  Additionally, the relationship between $\alpha$ and RS in Eq.~\ref{eq:rsfit} shows that increasing $\alpha$ after a spike allows the model to exhibit an increase in RS during the refractory period.  We choose the dynamics of these parameters to match the experimentally measured values reported in \cite{Miller2001}.  This study used a  paired pulse stimulation paradigm to probe the changes in threshold and RS due to spike time. Following an initial ``masker'' pulse that is sufficiently strong to always evoke a spike, there is an interpulse interval before a second pulse is presented.  The current level of the second pulse is varied in order to measure the firing efficiency curve of the neuron within the refractory period.

\cite{Miller2001} defined the dependence of $\theta$ and RS on the interpulse interval, which we denote as $\Delta t$, with the following equations:
\begin{align}
\label{eq:thresholdhistory}
\theta(\Delta t) = \frac{\theta_0}{ 1 - e^{-(\Delta t-t_{\theta})/\tau_{\theta}}}.
\end{align}
and
\begin{align}
\label{eq:rshistory}
RS(\Delta t) = \frac{RS_0}{1 - e^{-(\Delta t-t_{RS})/\tau_{RS}}}.
\end{align}
Here, $\theta_0$ and $RS_0$ are baseline values obtained from single pulse responses.  The parameter $t_{\theta}$ represents the absolute refractory period during which no spikes can be produced.  We implement the absolute refractory period by holding the intensity function $\lambda(t|I,H)$ at zero until the elapsed time since the last spike exceeds $t_\theta$.  The time constant $\tau_{\theta}$  quantifies how quickly the effects of a previous spike fade, and is therefore a measure of the relative refractory period.
The parameters $t_{RS}$ and $\tau_{RS}$ play similar roles in defining the evolution of RS, as a function of the time since the last spike.

To simulate the model with refractory effects, we let $\Delta t$ in the above equations represent the time since the last spike.  We then update the values of $\kappa$ and $\alpha$ at the onset of each pulse, based on the time since the last spike.  These parameters are then kept at a constant value from the onset of the pulse until the onset of the next pulse in the stimulus, at which time their values are updated again to reflect the changed time since the last spike.  This simplification, as opposed to allowing the parameters to vary continuously, allows a straightforward parameterization of $\kappa$ and $\alpha$ using the same single pulse relationships with $\theta$ and RS, respectively, which were derived above.  
The value of $\alpha$ must be obtained using numerical methods, but a formula for the value of $\kappa$ can be obtained by combining Eq.~\ref{eq:kappafit} and Eq.~\ref{eq:thresholdhistory}:
\begin{align}
\label{eq:kappahistory}
\kappa(\Delta t) = \kappa_0 ( 1 - e^{-(\Delta t-t_{\theta})/\tau_{\theta}}),
\end{align}
where $\kappa_0$ is the baseline value that is determined from the single pulse threshold $\theta_0$.
 
\subsection{Summary of parameterization method}
\label{subsec:summary}

The fitting sequence we have described above provides a semi-analytical method by which model parameters in Eqs.~\ref{eq:model} can be uniquely determined based on single and paired pulse response statistics.  This sequence can be summarized as follows:
\begin{enumerate}
\item Use RS value to determine nonlinearity parameter $\alpha$ in Eq.~\ref{eq:model1}.
\item Use $\alpha$ and chronaxie value to determine the stimulus filter time constant $\tau_\kappa$ in Eqs.~\ref{eq:model2} and~\ref{eq:model3}.
\item Use $\alpha$, $\tau_\kappa$ and the summation time constant to determine the value $\beta$ in Eq.~\ref{eq:model3} for scaling negative phases of the stimuli.
\item Use $\alpha$, $\tau_\kappa$, $\beta$, and threshold value to determine the scale factor $\kappa$ in Eqs.~\ref{eq:model2} and~\ref{eq:model3}.
\item Use $\alpha$, $\tau_\kappa$, $\beta$, $\kappa$ and jitter value to determine the jitter time constant $\tau_J$ in Eq.~\ref{eq:model4}.
\item Use the refractory function for RS in Eq.~\ref{eq:rshistory} and the relationship between RS and $\alpha$ to determine the dynamics of $\alpha$ after a spike.
\item Use the refractory function for threshold in Eq.~\ref{eq:thresholdhistory} and the relationship between threshold and $\kappa$ to determine the dynamics of $\kappa$ after a spike.
\end{enumerate}

Several comments are in order regarding the effect of pulse shape and duration on these parameters.  Threshold should decrease with increasing pulse duration \citep{Vandenhonert1984}, this property is captured for monophasic pulses by the chronaxie statistics and $\tau_\kappa$.  However, thresholds also depend on pulse shape.  In particular, thresholds measured with biphasic pulses are higher than those measured with monophasic pulse \citep{Miller2001bi}.  The model exhibits this property, but it is not quantitatively matched to experimental data.   In contrast, the unique relationship between RS and $\alpha$ in Eq.~\ref{eq:rsfit} shows that, for this model, RS does not depend on pulse shape or pulse duration.  This feature is consistent with data reported in \citep{Miller1999}, although others have suggested that RS may increase with pulse duration \citep{Bruce1999intensity}.  The parameters $\beta$ and $\tau_J$ will both depend on the pulse shape and durations in ways that we do not attempt to fit to physiological data.  Ideally one would define the model with a set of data obtaining using a self-consistent set of stimulation parameters, and these same stimulation parameters would then be used to obtain additional physiological recordings or perform cochlear implant psychophysics experiments.  We fit model parameters and compared subsequent simulation results to data from a variety of published sources; we will comment on variations in stimulation protocols and how they may impact our modeling results where appropriate.

\subsection{Numerical methods}
\label{subsec:numerical}

All simulations were performed using original computer code written in the Fortran programming language.  Numerical quadrature to evaluate the stimulus and jitter filters was performed using the trapezoid method and a time step of 1 $\mu s$.  Random numbers were generated using the Mersenne twister algorithm \citep{mt19937:FortranVer}.  The original Fortran code is available from the authors upon request and a more user-friendly version of the code written for Matlab (Mathworks, Inc.) can be downloaded from the ModelDB website (accession number 143760) or the website of one of the authors (http://www.cns.nyu.edu/$\sim$goldwyn/).

\section{Results}
\label{sec:result}

\subsection{Connections to biophysical mechanisms and models}
\label{subsec:biophysical}

In order to gain intuition into the structure and behavior of the point process model, we take a brief detour to explore its key features and parameters.  To provide this intuition, we connect aspects of the point process model to biophysical mechanisms and more familiar mathematical models of neurons.

\begin{figure}[t!]  
\includegraphics[width=84mm]{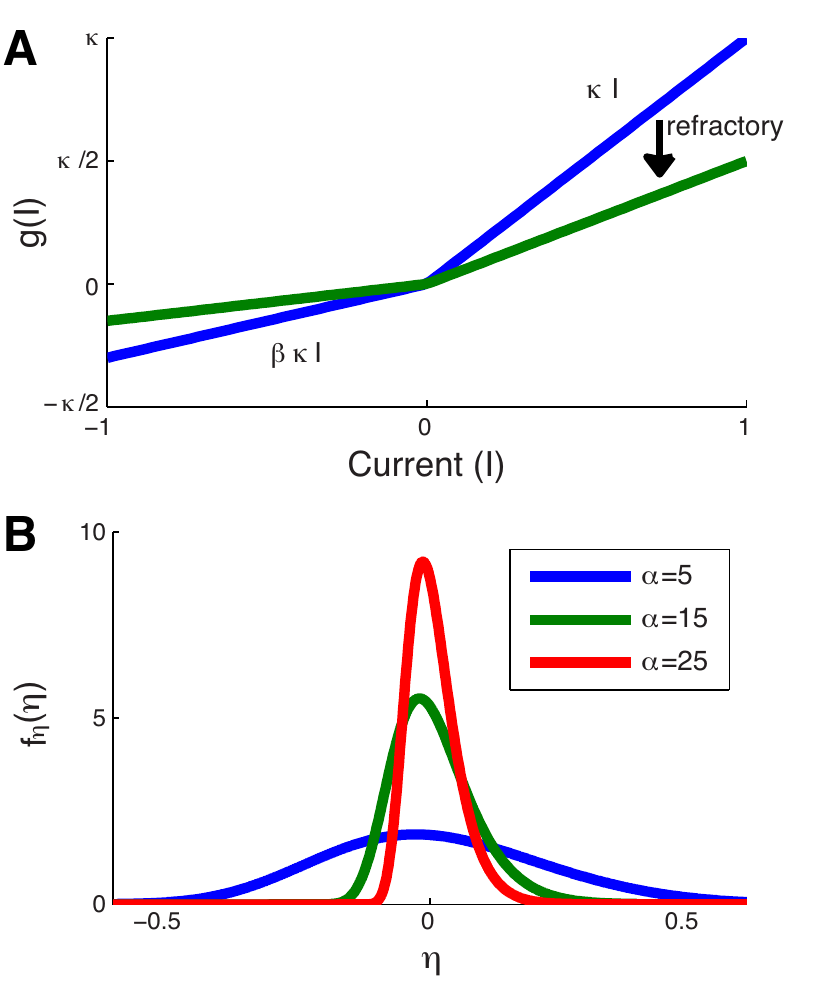}
\caption{ Relationship between the point process model and a soft-threshold integrate and fire model.  
\textbf{A:} Illustration of the function $g(I(t))$ in Eq.~\ref{eq:if}. The parameter $\beta$ controls the summation time constant of the model by decreasing the slope of $g(I(t))$ for negative currents.  Refractory effects reduce the excitability of the neuron by decreasing the slope for all current values.
\textbf{B:} Illustration of the probability density function for the noisy variable $\eta$ in the stochastic threshold analogy (see text).  Smaller values of $\alpha$ generate broader distributions, or equivalently a more variable spike initiation mechanism.}
\label{fig:ConnectionToIF}
\end{figure}

\subsubsection{Subthreshold dynamics} 
\label{subsubsec:subthreshold}
The result of applying the exponential filters $K^+(t)$ and $K^-(t)$ to the incoming stimulus $I(t)$ can be equated with a dynamic state variable that we denote by $v(t)$.  This notation is meant to suggest, in a loose sense, that this state variable represents the membrane potential of the model neuron.  We can reformulate the action of the stimulus filters $K^+$ and $K^-$ in terms of a differential equation, and find that the state variable $v(t)$ evolves according to:
\begin{align}
\label{eq:if}
\tau_\kappa \frac{dv}{dt} = -v + g(I(t)),
\end{align}
where the function $g(I)$ is pictured in Fig.~\ref{fig:ConnectionToIF}A and defined as:
\begin{align}
g(I) &=  \left\{  \begin{array}{l} \kappa(\Delta t) I \quad \mbox{if }I \ge 0  \\  \beta \kappa(\Delta t) I \qquad \mbox{else}  \end{array} \right. . \notag
\end{align}
The function $\kappa(\Delta t)$, which depends on the time $\Delta t$ since the last spike, is defined in Eq.~\ref{eq:kappahistory}.

The form of the differential equation for $v(t)$ in Eq.~\ref{eq:if} reveals that its dynamics are similar to the widely-used leaky integrate and fire model \citep{Burkitt2006a} and related spike response models \citep{Gerstner2002}, a fact noted in previous presentations of GLMs \citep{Paninski2004pp, Paninski2010}.  The effect of spike history on $g(I)$, which is determined by the dynamics of $\kappa(\Delta t)$ in Eq.~\ref{eq:kappahistory}, is to reduce the slope of this piecewise linear function, as shown in the transition from the blue to green lines in Fig.~\ref{fig:ConnectionToIF}A.  This change in $g(I)$ makes the model neuron less excitable immediately after a previous spike.  This feature has also been included in spike response models \citep{Gerstner2002} and integrate and fire models \citep{Badel2008}. 
The asymmetry in $g(I)$ enables the model to exhibit facilitation, or summation of consecutive charge-balanced pulses.  A modeling study using Hodgkin-Huxley type models has shown that summation of pseudomonophasic pulses, and in particular the time constant $\tau_{sum}$ in Eq.~\ref{eq:sumratio1}, is determined by the dynamics of the $m$-gating variable \citep{Cartee20002}.  The asymmetry in $g(I)$, therefore, can be viewed as a phenomenolgical approximation of the nonlinear process -- mediated by Na$^+$ activation -- by which the excitability of a neuron in response to consecutive charge-balanced pulses is increased beyond what would be expected if both pulses were presented in isolation from one another.

The parameter $\tau_\kappa$ appears in Eq.~\ref{eq:if} as the time scale of the dynamics of $v(t)$.  In the context of an integrate and fire model of a point neuron, this time scale is often interpreted as the membrane time constant and related to the passive integration properties of the neural membrane.  In the context of cochlear implants, such a description overlooks the fact that intracochlear electric fields can interact with AN fibers at multiple, anatomically diverse regions of the neural membrane.  For instance, membrane time constants of excitable nodes of Ranvier, somata, and unmyelinated portions of neurons may differ significantly~\citep{Rattay2001a}.  The parameter $\tau_\kappa$, therefore, must be interpreted as a single time scale that summarizes the combined integrative properties of a spatially extended AN fibers in an external electric field.

\subsubsection{Spike generation} 
\label{subsubsec:spike}
In the second stage of the model, the power law nonlinearity in Eq.~\ref{eq:model1} is applied to the state variable $v(t)$ to generate the desired probability of spiking.  In the GLM and related linear-nonlinear frameworks, this nonlinearity is typically static and often interpreted as a reduced description of the spike generating mechanisms of a neuron.  We have introduced the innovation that the nonlinearity depends on the time since the last spike in order to account for the observation that RS increases during the refractory period \citep{Miller2001}. Here, we explain the relationship between the nonlinearity and spike initiation and show why a dynamic nonlinearity produces the desired change in RS.

We begin by viewing the state variable $v(t)$ along with the nonlinear function $f(v) = v^\alpha$ as a definition of a point process conditional intensity function.  The probability of observing a spike in a small time window $\delta t$ can then be obtained by approximating the lifetime distribution function in Eq.~\ref{eq:Lifetime}.  We find that:
\begin{align}
\label{eq:lifetimedt}
\mbox{P(spike; } \delta t) \approx 1-e^{-v^\alpha \delta t}.
\end{align}

We now consider an alternate point of view in which $v(t)$ represents the subthreshold state of a neuron.  We suppose that the probability that a spike occurs in the $\delta t$ time window is related to the distance between $v(t)$ and some {\it voltage} threshold, which we denote $\Theta$.  This idea is known as an \emph{escape noise} model \citep{Plesser2000} and is closely related to stochastic threshold crossing models \citep{Bruce1999intensity, Litvak2003sinusoid}.  From this perspective, we can imagine that the noise free trajectory $v(t)$ is modified by adding a random number $\eta$ that represents the noise in the spike generation process.  The probability that a spike occurs in a $\delta t$ window can be written as:
\begin{align}
\label{eq:Pspike}
\mbox{P(spike; } \delta t) \approx \mbox{P}(v+ \eta >\Theta) . 
\end{align}
If we define the cumulative distribution function $F_\eta$ for the noise variable $\eta$, then we can equate Eqs.~\ref{eq:lifetimedt} and~\ref{eq:Pspike} and get:
\begin{align}
F_\eta(\eta) = e^{-(\Theta-\eta)^\alpha \delta t}  \notag
\end{align}
By taking the derivative of this distribution function, we derive a probability density function for $\eta$:
\begin{align}
f_\eta(\eta) = \alpha \delta t (\Theta-\eta)^{\alpha-1} e^{-(\Theta-\eta)^\alpha \delta t}. \notag
\end{align}

In Fig.~\ref{fig:ConnectionToIF}B, we illustrate the relationship between $\alpha$ and the probability density function of $\eta$ by plotting $f_\eta(\eta)$ for several values of $\alpha$.  We computed these distributions using $\delta t = 1 \mu s$ and $\Theta=\sqrt[\alpha]{\frac{\log 2}{\delta t }}$. The probability density function for $\eta$ becomes broad for small values of $\alpha$. This shows how the parameter $\alpha$ characterizes variability in the spike initiation process.
Moreover, this illustrates that allowing $\alpha$ to decrease immediately after a spike creates the desired behavior in the model -- a spike generation mechanism that is more variable within the refractory period.
To provide a biophysical interpretation for this parameter, we note that the random open and closing of ion channels in the cell membrane (known as channel noise) has been shown to determine the value of RS in simulation studies~\citep{Rubinstein1995}.  Thus, we view $\eta$ as an abstract representation of channel noise.

\subsubsection{Spike time variability} 
\label{subsubsec:timevariability}
The final stage of the model is to apply the jitter filter $J(t)$ to the output of the nonlinearity.  As we will see below, this secondary filter is necessary in order for this point process model to have realistic amounts of spike time variability.  $J(t)$ is applied after the nonlinearity, which suggests that it represents a source of timing variability subsequent to the spike generator in the cell.   In the context of extracellular electrical stimulation of the AN, one possible source of this variability is spatial interaction among multiple possible spatial sites of spike initiation.  For instance, in a simulation study of a spatially extended neuron with multiple nodes of Ranvier, \cite{Mino2004} showed that stimulating electrodes which are more distant from a model AN fiber produce more variable spike times. The more distant electric fields broaden the distribution of nodes of Ranvier at which spike initiation can occur, thus spatial interactions across the axon may represent one post-spike mechanism that adds jitter to spike times.

\subsection{Model parameterization}
\label{subsec:resultmodelparam}

\begin{table*}[t]
\caption{ Neural response statistics and corresponding model parameter values}
 \label{table:parameter}
  \begin{tabular}{lllll}  
    \hline \hline \noalign{\smallskip}
    Response statistic & Value & Reference & Model Parameter & Value \\ 
    \noalign{\smallskip}\hline\noalign{\smallskip}
    Threshold & 0.852 mA & \citep{Miller2001bi}    & $\kappa$                    & 9.342 \\ 
    Relative Spread & 4.87\%& \citep{Miller2001bi} & $\alpha$ & 24.52  \\ 
    Chronaxie & 276 $\mu$s& \citep{Vandenhonert1984}  & $\tau_\kappa$ & 325.4 $\mu s$ \\ 
    Jitter & 85.5 $\mu$s& \citep{Miller2001bi} & $\tau_J$ & 94.3 $\mu s$  \\
    Summation Time &  250 $\mu$s& \citep{Cartee2006} & $\beta$ &  0.333 \\
   \noalign{\smallskip}\hline
  \end{tabular}
\end{table*}

\begin{figure}[t!] 
\includegraphics[width=84mm]{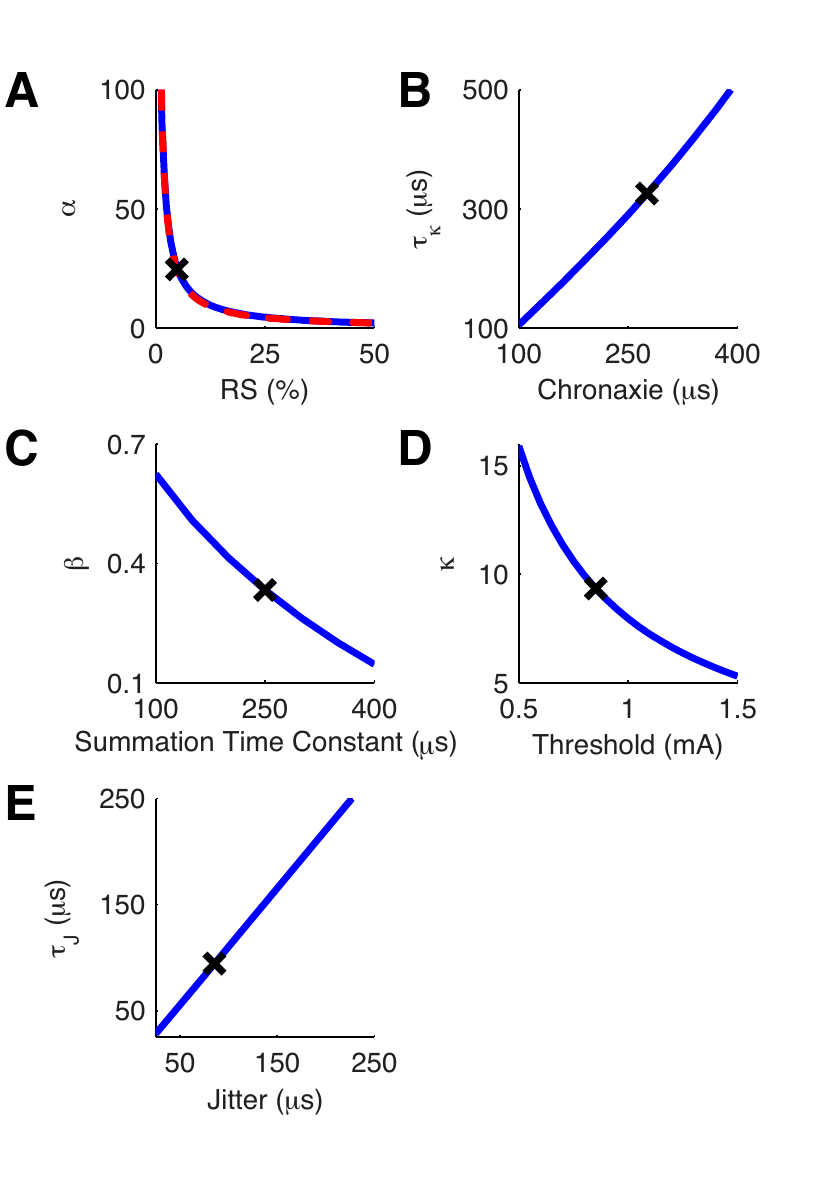} 
\caption{ Relationships between response statistics and model parameters.  Following the procedure summarized in Sec.~\ref{subsec:summary}, we obtain unique relationships between response statistics (x-axis) and model parameter values (y-axis).  At each step, a parameter value is chosen to fit the response statistic (black $\times$, see Table~\ref{table:parameter} for exact values), and then this parameter value can be used to obtain a unique relationship between the next response statistic and model parameter value in the fitting hierarchy.  The approximation used to determine $\alpha$ (Eq.~\ref{eq:alphaapprox}) is shown as a dashed red line in the first panel.}
\label{fig:param}
\end{figure}

With this understanding of the model features in hand, we proceed to parameterize the model following the steps outlined in Sec.~\ref{subsec:summary}. We defined parameters in the model using published data from single pulse and paired pulse recordings in cat.  Table~\ref{table:parameter} summarizes the data values we used, their sources in the literature, and the parameter values that we obtained from the fitting procedure.  Whenever possible, we used mean values obtained from experiments that used biphasic pulses: these include the mean threshold, RS, and jitter values reported in \citep{Miller2001bi}.  For chronaxie, we used the mean value reported by \cite{Vandenhonert1984} for intracochlear electrical stimulation of cat AN fibers. The longest phase duration tested was 2000 $\mu s$.  For the summation time constant ($\tau_{sum}$ in Eq.~\ref{eq:sumratio1}), we did not find a mean value reported in the published literature, so we assumed a value that fell within the range reported in \cite{Cartee2006} for AN responses to an electrode placed in the scala tympani.  The parameter values that describe the refractory effects on threshold and RS were obtained from \cite{Miller2001}.  Specifically, we used $t_\theta=332 \mu s$ and $\tau_\theta=411 \mu s$, which were the mean values reported in this study.   Mean values were not reported for $t_{RS}$ and $\tau_{RS}$, so we estimated that $t_{RS} = 199 \mu s$ and $\tau_{RS} = 423 \mu s$ based on plotted data (Figure 8 in \cite{Miller2001}).  These experiments used monophasic stimuli, but it is straightforward to modify our refractory model should data obtained using biphasic pulses become available.

Fig.~\ref{fig:param} depicts the relationships between the response measures shown on the x-axes and the model parameters shown on the y-axes.  Each panel illustrates the unique relationships that arose by following the parameterization sequence summarized in Sec.~\ref{subsec:summary}.  For instance, in panel A we show the dependence of the nonlinearity parameter $\alpha$ on RS.  Once we had fixed the value of $\alpha$ based on the desired RS value, we then computed how the stimulus filter time constant $\tau_\kappa$ depended on chronaxie.  This relationship is shown in panel B.  One-by-one, we obtained values for each parameter, the values of which are marked by the $\times$ symbol in Fig.~\ref{fig:param} and reported in Table~\ref{table:parameter}.
 
In order to introduce spike history effects on RS, we must recompute $\alpha$ at the onset of every pulse.  In principle, we could use a numerical root finding method to invert Eq.~\ref{eq:rsfit}, but to avoid this and thereby increase the computational speed of our simulations, we observed that the relationship between $\alpha$ and RS could be approximated by the power law:
\begin{align}
\label{eq:alphaapprox}
\alpha \approx  {RS}^{-1.0587}.
\end{align}
The combination of this equation for $\alpha$ and Eq.~\ref{eq:rshistory} provided a simple rule for evolving $\alpha$ according to the time since the last spike.
The approximation, which is shown with the red dashed line in the first panel of Fig.~\ref{fig:param}, is suitably accurate.

\subsection{Model predictions: single pulse stimuli}
\label{subsec:predictsinglepulse}

\begin{figure}[t!]
\includegraphics[width=84mm]{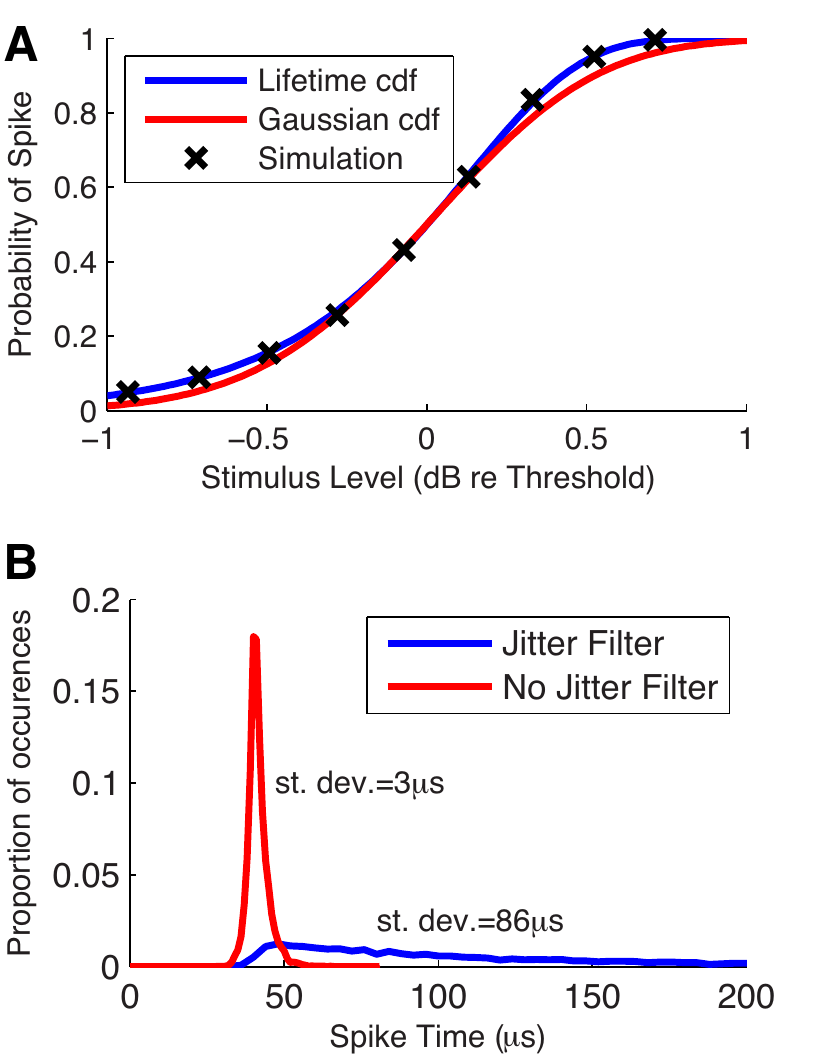} 
\caption{ Responses of the point process model to single pulse stimulation.
\textbf{A:} Firing efficiency curves for the point process model (blue), an integrated Gaussian function with the same mean and RS (red), and simulation results from the model (black $\times$).
\textbf{B:} Distributions of spike times obtained from 5000 simulations of the model with a jitter filter (blue) and without a jitter filter (red).}
\label{fig:SinglePulse}
\end{figure}

\subsubsection{Firing efficiency curve}
\label{subsubsec:fe}
We first simulated responses of the model to a single pulse of current.  To be consistent with the experimental data from which we obtained values for threshold, RS, and jitter, we used a 40~$\mu$s per phase biphasic pulse stimulus as the input to the model.  By varying the current level of the pulse, we could measure the complete firing efficiency curve, as shown in Fig.~\ref{fig:SinglePulse}A.  The probability of a spike is defined as the proportion of trials, out of a total of 5000 for each current level, on which the model generated a spike.  Simulation results are shown with black $\times$-marks.  As expected, they line up with the analytically derived firing efficiency curve obtained from Eq.~\ref{eq:Lifetime} and plotted with a blue line.  In order to compare this input/output function to the more standard functional form that is used in the stochastic threshold crossing model of \cite{Bruce1999single}, we show in red the integral of a Gaussian distribution with mean and standard deviation consistent with the threshold and relative spread values that we used to fit the point process model.  At the highest and lowest current levels, the firing efficiency curve for the point process model is slightly greater, but overall both methods produce similar spiking probabilities in response to a single pulse. 

\subsubsection{Jitter}
\label{subsubsec:resultjitter}
Fig.~\ref{fig:SinglePulse}B shows the distribution of 10000 spike times obtained from the point process model in response to a single biphasic pulse of phase duration 40~$\mu s$ presented at the threshold stimulus level.  The blue line shows shows results for the model with a jitter filter time constant of $\tau_J=94.3 \mu s$, as determined by the parameterization method.  This produces a distribution of spike times with standard deviation of $86 \mu s$, consistent with the mean value in \citep{Miller2001bi} that was used to parameterize the model.  In the absence of a second filtering stage, the model produces an extremely narrow distribution of spike times, as shown by the red distribution.  The secondary filter is necessary, therefore, for the model to have realistic amounts of spike time jitter.

\subsection{Model predictions: paired pulse stimuli}
\label{subsec:predictpairedpulse}

\subsubsection{Summation pulse paradigm}
\label{subsubsec:summation}

\begin{figure}[t!]
\includegraphics[width=84mm]{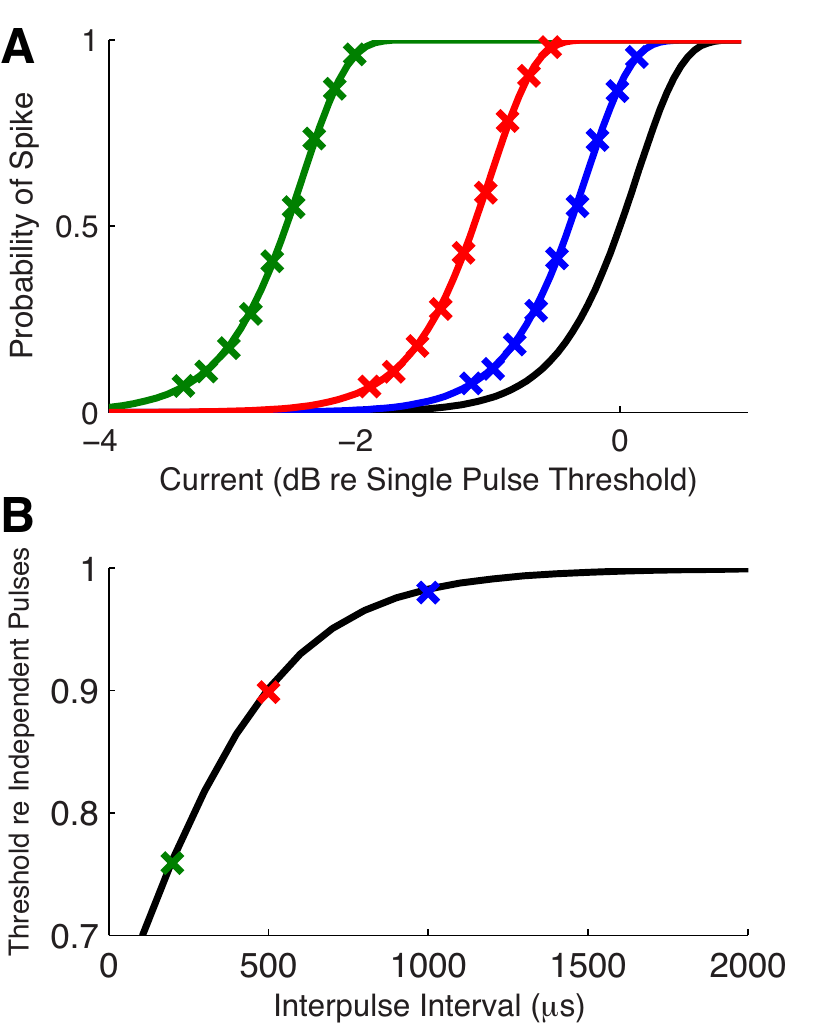} 
\caption{ Responses of the point process model to the summation pulse paradigm.
\textbf{A:} Simulated spiking probabilities (symbols) for three interpulse intervals:  200~$\mu s$ (green), 500~$\mu s$ (red), 1000~$\mu s$ (blue).  Solid lines are analytically obtained from the lifetime distribution function in Eq.~\ref{eq:Lifetime}.  Single pulse firing efficiency curve is shown in black for reference.
\textbf{B:} Thresholds estimated from simulated firing efficiency curves and normalized by the threshold for two independent pulses are shown as $\times$-marks for the three interpulse intervals.  Facilitation occurs if this threshold ratio is less than one. Black curve is the analytically predicted result in Eq.~\ref{eq:sumratio2}.  }
\label{fig:SummationPulse}
\end{figure}

To test how the neuron model responds to pairs of pulses, we first simulated the summation pulse procedure in \citep{Cartee2006}.  As inputs, we used two biphasic pulses of equal current level separated by an interpulse interval of either 200$\mu s$, 500$\mu s$, or 1$m s$.  The probabilities of observing at least one spike in response to both pulses, for varying interpulse intervals and current levels, are shown as $\times$-marks in Fig.~\ref{fig:SummationPulse}A.  These were computed from 5000 repeated simulations of the model.  We also plotted the analytically obtained firing probabilities for pulse pairs, obtained from Eq.~\ref{eq:Lifetime}, as curves in panel A.  The effect of summation is visible as a leftward shift of these curves at shorter interpulse intervals.  As a reference, the black line reproduces the single pulse firing efficiency curve from Fig.~\ref{fig:SinglePulse}A.

To summarize these results, we estimated summation thresholds at each interpulse interval by fitting an integrated Gaussian function.  We then normalized the estimated threshold values relative to the single pulse threshold and plotted the values in Fig.~\ref{fig:SummationPulse}B.  The black curve represents the analytical prediction of the point process model,  given by Eq.~\ref{eq:sumratio2}, for the threshold of two independent pulses.  Facilitation occurs if the threshold ratio is less than one.  In these simulations we see facilitation for interpulse intervals shorter than approximately 1000 $\mu s$.  The point process model, therefore, will exhibit facilitation for carrier pulse rates of roughly 1000~pps and above.

\subsubsection{Refractory pulse paradigm}
\label{subsubsec:refractorypulse}

\begin{figure}[t!]
\includegraphics[width=84mm]{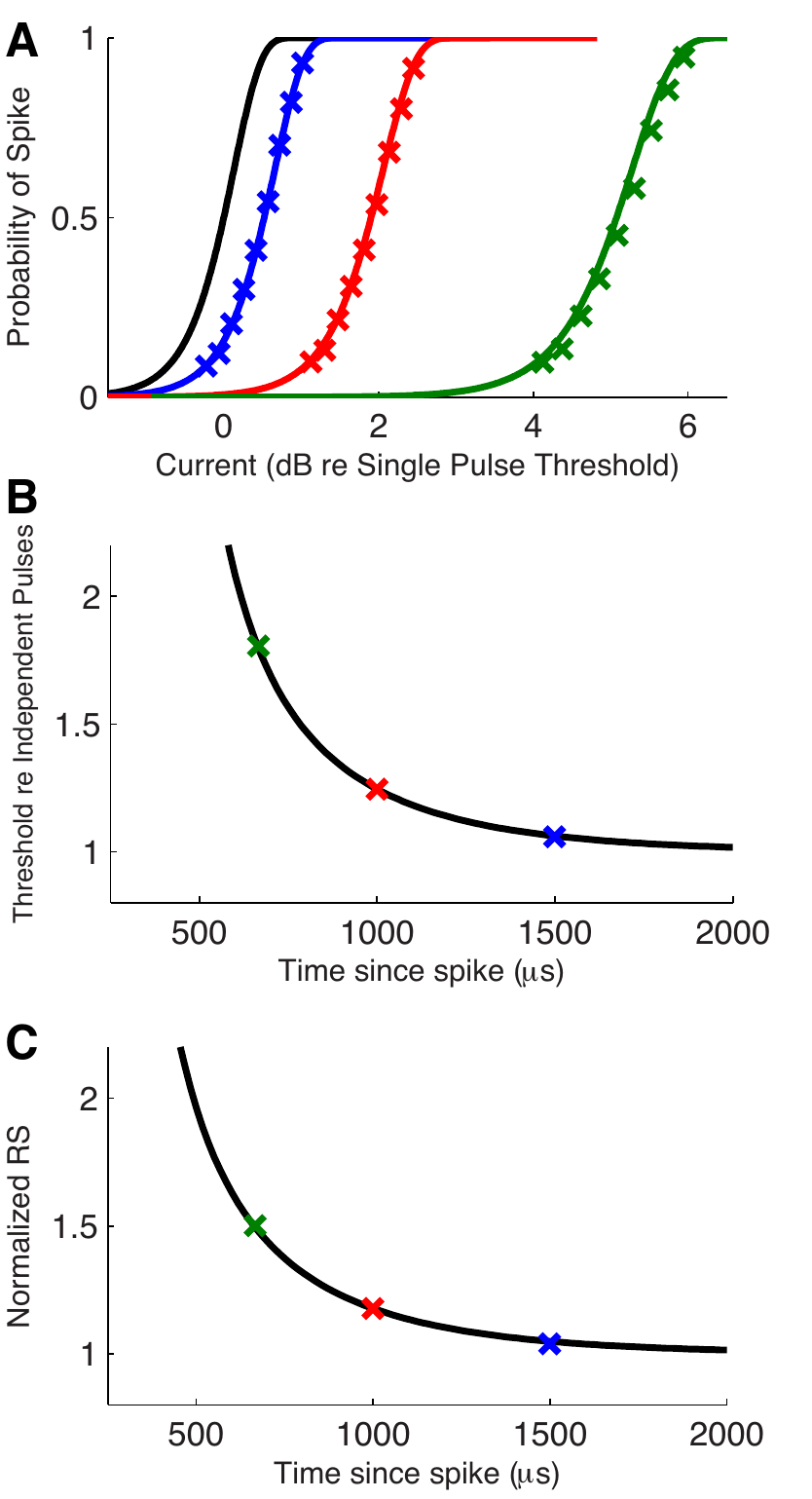} 
\caption{ Responses of the point process model to the refractory pulse paradigm.
\textbf{A:} Simulated spiking probabilities (symbols) for three interpulse intervals: 667~$\mu s$ (green), 1000~$\mu s$ (red), 1500~$\mu s$ (blue).  Solid lines are analytically obtained firing efficiency curves obtained from lifetime distribution function in Eq.~\ref{eq:Lifetime}.  Single pulse firing efficiency curve is shown in black for reference.
\textbf{B:} Thresholds estimated from the simulated firing efficiency curves and normalized by the single pulse threshold are shown as symbols for the three interpulse intervals.  Black curve is Eq.~\ref{eq:thresholdhistory}.
\textbf{C:} Relative spread estimated from the simulated firing efficiency curves and normalized by the single pulse relative spread are shown as symbols for the three interpulse intervals.  Black curve is Eq.~\ref{eq:rshistory}.}
\label{fig:RefractoryPulse}
\end{figure}

To probe the effects of spike history in the model, we followed the experimental procedure in \citep{Miller2001} and simulated responses to masker-probe pulse pairs. The current level of the first pulse was set to a very high value so that it always elicited a spike.  The level of the second pulse was then varied in order to measure firing efficiency curves.  We used pairs of biphasic pulses in order to be consistent with our previous simulations.  Spike probabilities were defined as the proportion of trials (out of 5000 total) in which the second pulse produced a spike.  Firing efficiency curves obtained from responses to the second pulse of current are shown in Fig.~\ref{fig:RefractoryPulse}A for three different interpulse intervals (667, 1000, and 1500 $\mu s$).  As the interpulse interval (and equivalently the time since the last spike) decreases, the spike history effect has a greater influence on the probability that the second pulse will evoke a spike.  

As discussed in Sec.~\ref{subsubsec:Refractory}, there are two types of refractory effects in the model.  The first is the standard refractory phenomenon whereby the model neuron is less excitable immediately after a spike.  This is apparent in the rightward shift in the firing efficiency curves for shorter interpulse intervals in Fig.~\ref{fig:RefractoryPulse}A.  The second effect is that RS increases at shorter interpulse intervals. This leads to shallower slopes in the firing efficiency curves at shorter interpulse intervals, but the decibel scale in panel A obscures the change. 
There are some discrepancies between the simulation results ($\times$-marks) and the predicted values from point process theory at the smallest interpulse intervals (green line). The differences arise from the fact that the simulated spike times in response to the first pulse have a small amount of random variability, whereas the analytical result is computed using the assumption that the time of the first spike is locked to the time of the onset of the first pulse.  

To summarize these effects, we estimated threshold and RS from the simulated firing efficiency curves by fitting integrated Gaussian functions.  These values are normalized with respect to single pulse threshold and RS and shown in panels B and C, respectively.  By construction of the model, the values computed from simulations agree with the refractory functions Eq.~\ref{eq:thresholdhistory} and Eq.~\ref{eq:rshistory}, which we plot as black curves.  We emphasize that a model with a static nonlinearity, namely one in which $\alpha$ does not depend on spike history, would produce RS values that decrease at shorter interpulse intervals.  Alternatively, the stochastic threshold crossing model of \cite{Bruce1999train} assumes a constant RS and would produce a flat line in panel C.

\subsection{Model predictions: constant pulse train stimuli}
\label{subsec:predictconstanttrain}

The results to this point have validated the fitting method and demonstrated that the model reproduces a range of measures that characterize responses to single pulses and pairs of pulses.  Of greater relevance to cochlear implant speech processing strategies are the responses of neurons to extended trains of pulsatile stimuli.  We first simulated responses to trains of biphasic pulses with phase duration 40~$ \mu s$ and constant current levels, and sought to characterize how carrier pulse rate affects the sequence of evoked spikes.  We relate our simulation results to relevant physiological data throughout.
  
\subsubsection{Firing rate and interspike interval distributions }
\label{subsubsec:firingrate}

\begin{figure*}[t!]
\includegraphics[width=174mm]{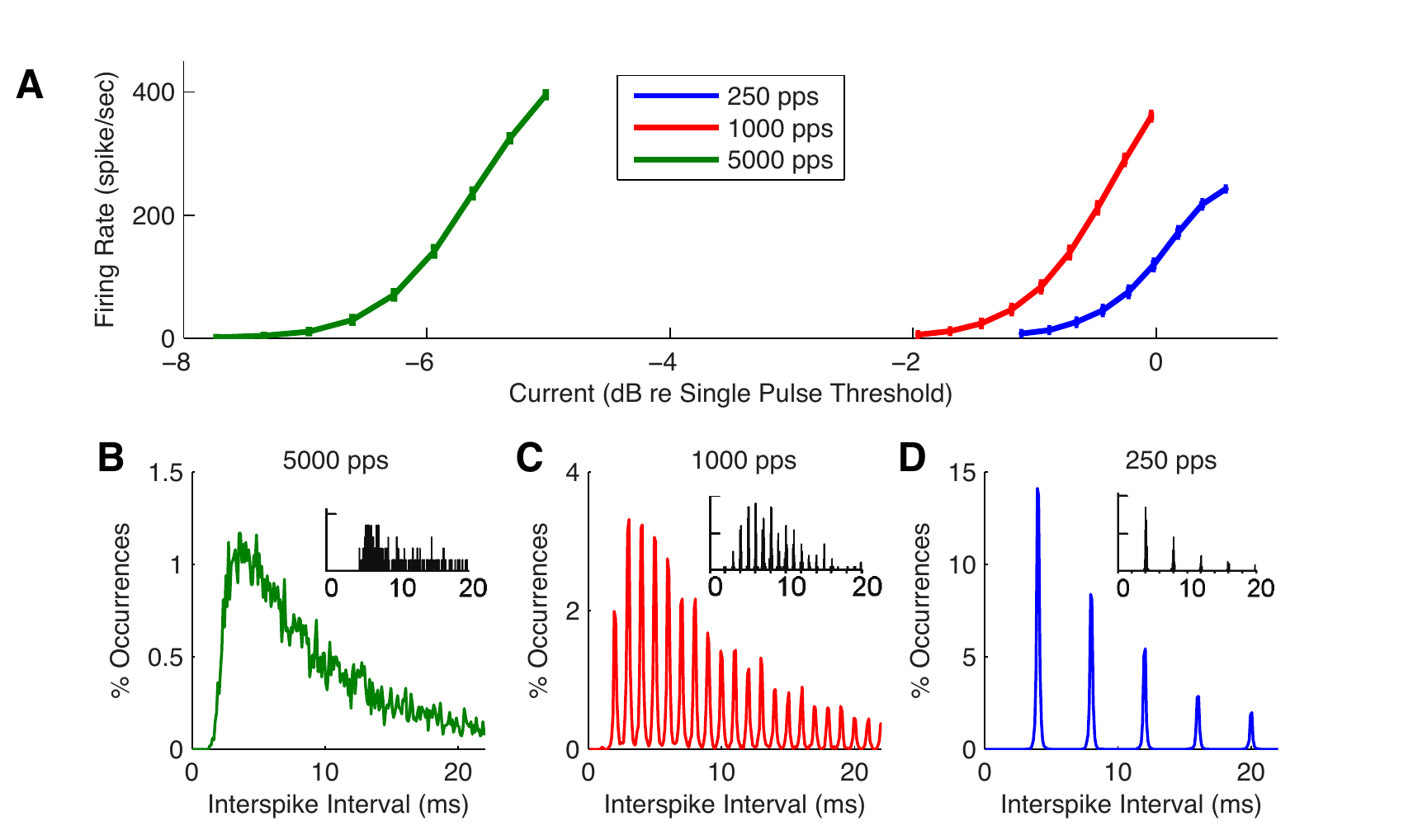} 
\caption{ Firing rate and interspike interval distributions for three stimulus pulse rates.
\textbf{A:} Firing rate as a function of current level per pulse.  Stimuli were one second long trains of biphasic pulses with constant current level per pulse, the level of which is shown on the x-axis.  Error bars indicate the standard deviation of the mean firing rate, estimated from 100 repeated simulations of the model.
\textbf{B-D:} Interspike interval distributions estimated from 10000 interspike intervals with the pulse rates indicated in each panel.  Current level per pulse was chosen so that the stimuli evoked approximately 100 spikes per second.  Inset figures are single fiber recordings from cat AN fibers at the corresponding pulse rates reproduced from Fig. 1 in \cite{Miller2008changes} with kind permission from Springer Science+Business Media: J. Assoc. Res. Otolaryngol.}
\label{fig:RateAndISI}
\end{figure*}

Fig.~\ref{fig:RateAndISI}A shows how firing rates in the model increase with current level for three different pulse rates.  The stimuli were one second in duration. Mean and standard deviations (shown with error bars) were estimated from 100 repeated simulations.  At the lowest pulse rate (250~pps, blue line), the interpulse interval is longer than the  summation time scale as well as the relative refractory periods for threshold and RS.  Thus the firing rate curve follows directly from the single pulse firing efficiency curve in Fig.~\ref{fig:SinglePulse}A.  The firing rate saturates at one spike per pulse, in this case 250 spikes per second.  The 1000~pps and 5000~pps pulse trains have interpulse intervals of 1 ms and 200 $\mu s$, respectively, which are short enough for refractory and summation effects begin to impact the behavior of the model.  Due to summation and the higher pulse rate, the same firing rate can be evoked by higher pulse rate stimuli using less current per pulse. Thus the firing rate curves are shifted leftward as the pulse rate increases.

The remaining panels of Fig.~\ref{fig:RateAndISI} show interspike interval distributions at the three pulse rates, where the current levels were set to evoke approximately 100 spikes per second.  The distributions were obtained from histograms of 10000 spike times in time bins of 100 $\mu s$.  The distributions obtained from responses to the 250~pps and 1000~pps pulse trains, but not the high rate 5000~pps pulse train, show peaks that are clearly aligned to the interpulse intervals in the stimulus.  This feature of the simulated interspike interval distributions is qualitatively similar to distributions recorded from cat AN fibers using the same pulse rates \citep{Miller2008changes}.  For ease of comparison, we include examples of interspike interval data reported by Miller et al in the insets of each panel.  The model is capable of reproducing some characteristics of the interspike interval distribution for this single neuron, although an important caveat is that the experimentally-measured interspike intervals were obtained under different stimulus conditions and likely also different firing rates.  We do not intend to claim that the model completely reproduces all of the behavior of this or other cells.

The results of these simulations suggest that spike time jitter, which was incorporated into the model on the basis of responses to single pulse stimuli, can account for distinctive features of the interspike interval distributions and how they vary with the stimulus pulse rate.  On the one hand, temporal variability in spike times is small relative to the interpulse intervals for the lower pulse rate stimuli, which leads to the periodic nature of these distributions in Fig.~\ref{fig:RateAndISI}C and D.  On the other hand, this small temporal variability is sufficient to spread the simulated spike times across the interpulse interval for the 5000~pps, creating a more smoothly varying interspike interval distribution in panel B, consistent with the appearance of the distribution shown in the inset.  

\subsubsection{Synchronization of spike times}
\label{subsubsec:synch}

\begin{figure}[t!]
\includegraphics[width=84mm]{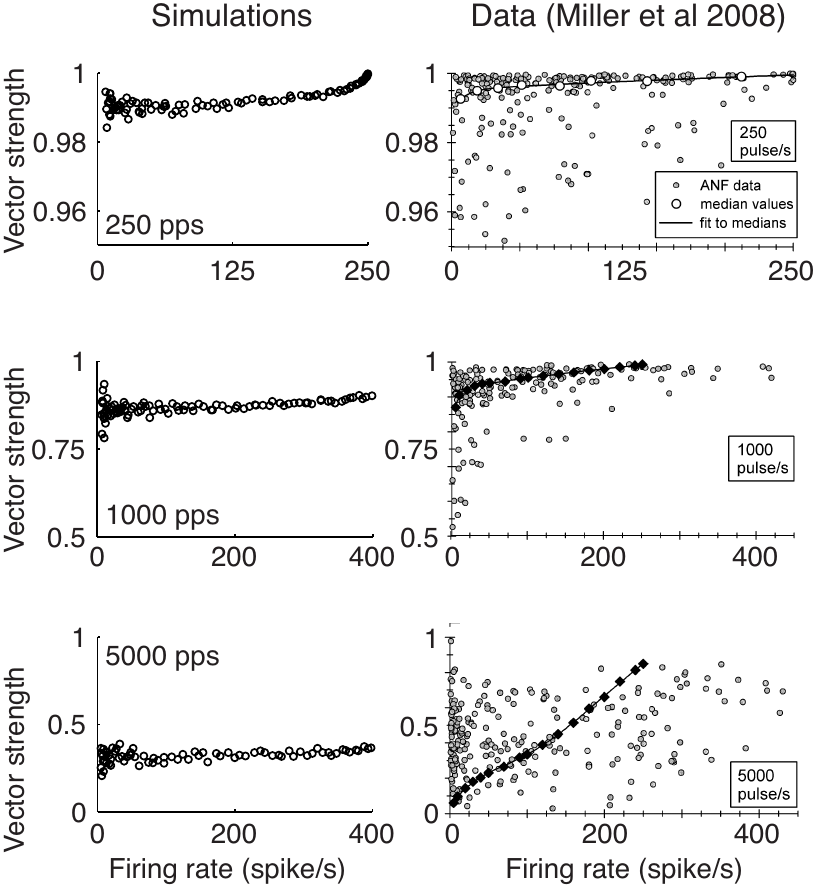} 
\caption{ Vector strength measured from responses to three pulse rates.
\textbf{Left column:} VS computed from simulated responses to 10 second long pulse trains of biphasic pulses with equal current levels.  Current levels were varied to explore a range of firing rates.  Note the change of scale on the y-axis; VS values decrease as pulse rate increases.
\textbf{Right column}: VS values reported by \cite{Miller2008changes} based on responses of 37 cat AN fibers stimulated with intracochlear electrodes using monopolar, biphasic pulse trains. Reproduced from Fig. 8 in \citep{Miller2008changes} with kind permission from Springer Science+Business Media: J. Assoc. Res. Otolaryngol. The black diamonds do not represent median values, see \cite{Miller2008changes} for details.}
\label{fig:VS}
\end{figure}

Simulation and physiological studies have generated interest in the possibility that high rate pulse trains may desynchronize neural responses to cochlear implant stimulation \citep{Rubinstein1999, Litvak2003desynch}.  We tested whether the model exhibited similar signatures of desynchronization by computing a synchronization index, known as vector strength \citep{Goldberg1969}, and then compared our results to physiological data reported in \cite{Miller2008changes}.   For a sequence of $N$ spike times $\{ t_i \}_{1}^N$, vector strength is defined with respect to a period $T$ as:
\begin{align}
\label{eq:VS}
 \mathrm{VS} = \frac{1}{N}\sqrt{ \left[\sum_{i=1}^N \cos (2 \pi t_i/T) \right]^2+ \left[\sum_{i=1}^N\sin (2 \pi t_i/T) \right]^2}.
\end{align}
VS takes values between 0 and 1, with higher values being interpreted as a more synchronized spike train.  In these simulations, we use VS to measure the strength of phase locking to the period of the stimulus pulse train, and thus $T$ represents the interpulse interval.  In all simulations, as well as the experimental data to which we compare our results, the stimulus is a train of biphasic pulses with a constant current level and a 40~$\mu s$ phase duration.  Simulation results for three pulse rates are shown in the left column of Fig.~\ref{fig:VS}, where each circle represents the firing rate and VS values computed from the response to a 10 second long pulse train.  Our results can be compared to VS values obtained from recordings of cat AN fibers responding to biphasic pulses trains presented at the same pulse rates and pulse shape.  These data, which we reproduce from \cite{Miller2008changes}, are shown in the right column of Fig.~\ref{fig:VS}.

For the lowest pulse rate, VS values obtained from simulations exceeded 0.98 for all firing rates.  This represents near perfect phase locking to the 250~pps pulse train.  VS systematically decreases with pulse rate -- note the change of scale on the y-axes.  Comparisons with the data of \cite{Miller2008changes} show good agreement between simulated and measured VS values.  The primary discrepancy between the two sets of VS values is the large variability present in the data values, seen as scatter in the vertical direction of these plots.  A likely cause of this difference is the fact that we simulated a single model neuron with a fixed set of model parameters.  In contrast, the data were obtained from a sampling of 37 AN fibers. Presumably each neuron had different intrinsic properties that may even change further over the course of the experiment.  Additionally, Miller et al suggested that some scatter in the VS values measured from responses to the 5000~pps stimulus may have been due to limitations in their ability to resolve the phase of spike times with respect to the 200~$\mu s$ interpulse interval.

\begin{figure}[t!]
\includegraphics[width=84mm]{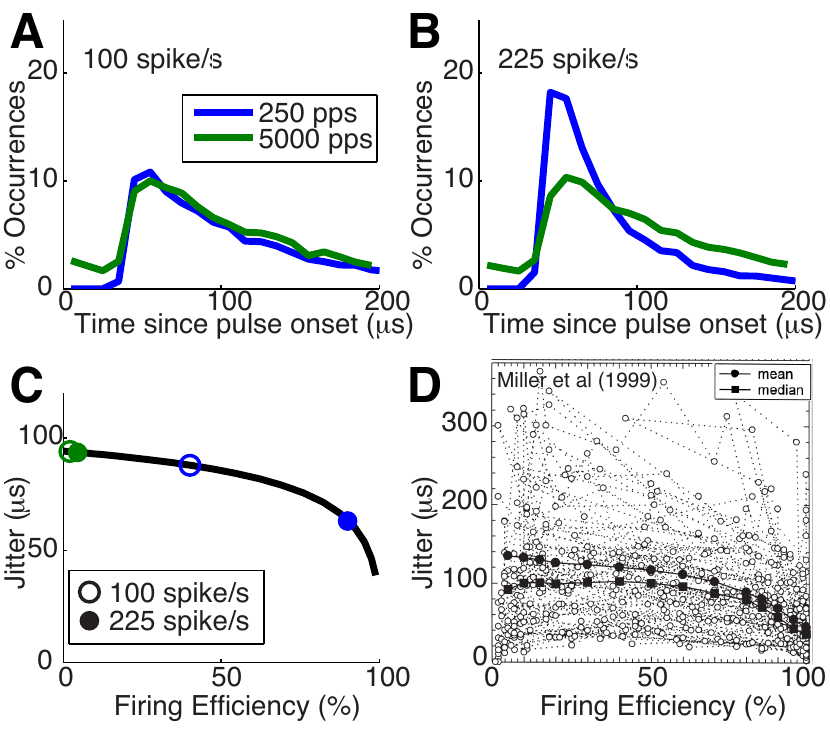} 
\caption{ Further comparison of spike timing variability in response to low and high rate pulse trains.
\textbf{A:} Distribution of times between simulated spikes and the onset of the previous pulse.  Stimuli are trains of biphasic pulses of either 250~pps (blue) or 5000~pps (green), with a constant  current level per pulse.  Current level was chosen so that firing rates of the model were approximately 100 spikes per second for both stimuli. 10000 simulated spike times were used to estimate the distributions, which were plotted using $10 \mu s$ bin window.
\textbf{B:} Similar to \textbf{A}, with current levels increased so that firing rates in simulations were approximately 250 spikes per second.
\textbf{C:} Similar relationship between jitter and firing efficiency in the point process model.  Black curve represents analytical calculation of jitter from lifetime distribution function, blue circles mark the firing efficiency values, interpreted as average probability of a spike per pulse, corresponding to the pulse rates and firing rates in \textbf{B} and \textbf{C}.
\textbf{D:} Relationship between and jitter and firing efficiency in measurements of responses of cat AN fibers to monopolar, monophasic (cathodic) stimulation by an intracochlear electrode.  Reprinted from Hearing Research \citep{Miller1999} with permission from Elsevier.
}
\label{fig:DesynchExplain}
\end{figure}

On the basis of the lower VS values at higher pulse rates, one is tempted to conclude that these results reveal the desynchronizing effects of high pulse rate stimulation.  As noted in \cite{Miller2008changes}, however, higher VS values at higher carrier pulse rates may not indicate desynchronization since VS is computed with a different reference period for each stimulus, depending on the interpulse interval.  In particular, at higher carrier pulse rates the period $T$ in Eq.~\ref{eq:VS} is smaller, and this can lead to lower VS values for spike trains with equivalent amounts of temporal dispersion.  To test whether the low VS values computed in simulations using 5000~pps pulse trains do in fact indicate desynchronizing effects of high pulse rate stimulation, we examined the distribution of spike times relative to the onset time of the pulse immediately preceding the spike.  These spike time distributions are shown in Fig.~\ref{fig:DesynchExplain} and were estimated from 10000 spike times using a $10\mu s$ bin width.  Panel A shows distributions for the 250~pps (blue) and 5000~pps (green) pulse trains for a current level set to evoke approximately 100 spikes per second.  In panel B, we show distributions of spike times obtained from stimuli that caused the model to spike at 225 spikes per second.  The spike time distributions are nearly identical at the lower firing rate indicating that the lower VS values obtained with the 5000~pps stimulus do not reveal any desynchronization in this case.  At the higher firing rate, however, the distribution of spike times in response to the 250~pps stimulus is considerably narrower than the distribution of spike times measured from the 5000~pps pulse train.  One could interpret this as desynchronization, although it may be more accurate to say that the 5000~pps stimulus is maintaining a degree of spike time variability that is lost when the current level of the 250~pps stimulus is increased and the evoked firing rate approaches the maximal value of 250 spikes per second.

We can explain the narrowing of the spike time distribution by considering the main source of spike time variability in the model.  In Fig.~\ref{fig:DesynchExplain}C, we show the amount of jitter in simulated spike times, as a function of firing efficiency for  single pulse stimulation.  We obtained this curve directly from the density function for spike times in Eq.~\ref{eq:jitterpdf}.
For the 250~pps stimulus, firing rates of 100 and 225 spike per second translate to firing efficiency values per pulse of 40$\%$ and 90$\%$, respectively.  The blue circles mark the location of these firing efficiency values.  Jitter in the model decreases substantially at the higher firing efficiency value, which leads to the narrower spike time distribution in Fig.~\ref{fig:DesynchExplain}B.  In contrast, increasing the firing rate from 100 to 225 spikes per second when using a 5000~pps stimulus translates to a small change in the firing efficiency, here interpreted as the average probability of a spike per pulse, from $2\%$ to $4.5\%$ (green circles).  The model predicts, therefore, that high rate pulse trains can maintain temporal variability in spike times even at relatively high firing rates because the probability of a spike on any single pulse remains low.   The key feature of the model that explains these results -- the fact that jitter decreases with firing efficiency -- has also been observed in the responses of cat AN fibers to electric stimulation using monophasic pulses \citep{Miller1999}.  These data are reproduced in  Fig.~\ref{fig:DesynchExplain}D. They show a qualitative match with our point process predictions, and underline the conclusion that the desynchronization observed at high pulse rates largely follows from the use of weaker current impulses.

\subsubsection{Irregularity of firing responses}
\label{subsubsec:irregularity}

\begin{figure}[t!]
\includegraphics[width=84mm]{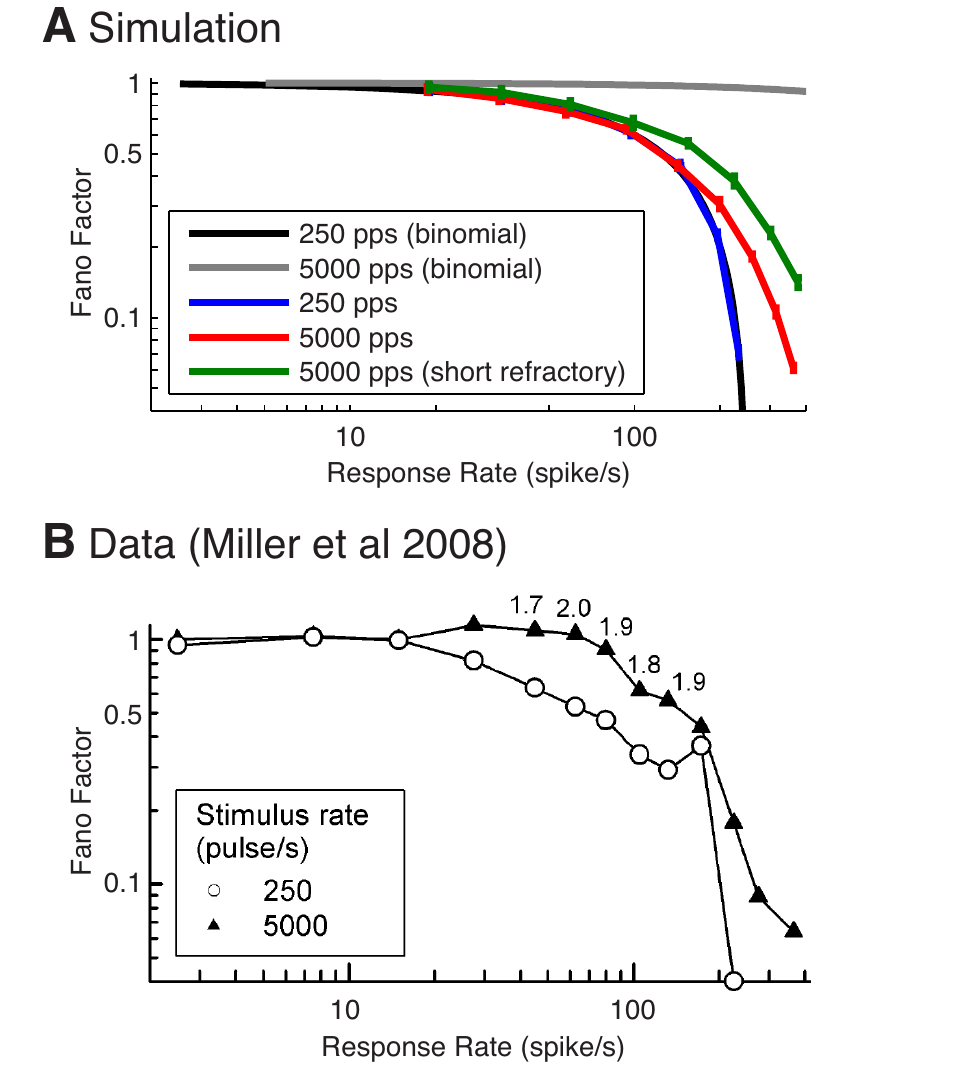} 
\caption{ Fano factor measured from responses to low (250~pps) and high (5000~pps) pulse rates.
\textbf{A:} Simulation results using the point process model.  Results for the low pulse rate stimulation (blue line) are predicted by the Fano factor for a binomial distribution (black line, see text for details).  The gray line shows the corresponding prediction of a binomial distribution for the 5000~pps pulse train.  The green line illustrates that simulated responses to the 5000~pps pulse train can have larger Fano factors than responses to the 250~pps pulse train if the refractory period is shortened ($\tau_\theta=200$~$\mu s$).  
Fano factor values are estimated from 100 intervals of simulated spike trains, where each interval is 100~s long.  Error bars represent standard error in the mean of these estimates.
\textbf{B:}  Medians of Fano factors recorded from cat auditory nerve fibers, reproduced from Fig. 9 in \citep{Miller2008changes} with kind permission from Springer Science+Business Media: J. Assoc. Res. Otolaryngol.}
\label{fig:Fano}
\end{figure}

An additional response statistic that has been used to characterize responses to high rate pulse trains is the Fano factor.  This quantity measures irregularity in the number of observed spikes.  It is defined as the variance in the number of spikes observed in a time window normalized by the mean value. 
A Fano factor of one, which is the value generated by any Poisson process model of spiking, is considered to signify a highly irregular spike generator and lower values indicate more regularity.  To estimate Fano factor from the model, we simulated responses to a one thousand second long train of biphasic pulses (40~$\mu s$ pulse duration), using 250~pps and 5000~pps pulse trains.  Ten estimates of Fano factor were obtained by subdividing these spike trains.  Results are shown in Fig.~\ref{fig:Fano}A, with error bars that represent standard error of the ten estimates of Fano factor for each stimulus condition.  In Fig.~\ref{fig:Fano}B, we reproduce a figure from \citep{Miller2008changes} in order to compare the behavior of the point process model to AN fibers in cat, also stimulated by biphasic pulse trains.  These are median values obtained from recordings of 37 AN fibers in 8 deafened cats, as such they do not reveal the considerable variation around the medians present in the experimental data. 

Fano factors obtained from simulations in response to 250~pps pulse trains are shown by the blue curve in Fig.~\ref{fig:Fano}A and results for the high rate pulse trains are in red.   The clear trend is a decrease in Fano factor as firing rate increases, consistent with the physiological data in panel B.  As firing rates approach 250 spikes per second, the responses to the low rate stimulus saturate at one spike for every pulse, leading to a sharp decrease in the Fano factor, a trend also seen in the data.  One substantial discrepancy between simulation results and the median Fano factors reported in \cite{Miller2008changes} is that, in the data, Fano factors obtained from responses to high pulse rate stimuli were  larger than those obtained from response to the low pulse rate stimuli for most firing rates.  The numbers above datapoints in Fig.~\ref{fig:Fano}B indicate the multiplicative factor by which the high rate Fano factors exceed the low rate values.  These experimental results supported the notion that high pulse rate stimulation can generate more irregular spiking activity.  

Can our model predict this phenomenon?   
As we show in greater detail in the Appendix, the effects of refractoriness and interpulse interactions tend to decrease Fano factor; see also~\citep{Berry1998}. 
When the response of a neuron to any pulse is independent of all previous stimuli and spike history, the activity can be described as a sequence of Bernoulli trials, where the probability $p$ that any one pulse evokes a spikes is equal to the average firing rate divided by the number of pulses.  The total number of spikes, in this case, would be distributed binomially with mean $Np$ and variance $Np(1-p)$, where $N$ is the total number of pulses.  If we denote the average firing rate by $r$ and the pulse rate by $\rho$, then the Fano factor would have a simple dependence on $p$, and, consequently, the average firing rate $r$:
\begin{align}
\label{eq:fbinomial}
F_{binomial} = 1-p = 1 - \frac{r}{\rho}. 
\end{align}

Our simulation results for the 250~pps pulse train are completely explained by this analogy to a binomial distribution.  Due to the relatively long interpulse interval of 4 ms, past spike and stimulus histories have no effect on the model's response to subsequent pulses.  The curve representing $F_{binomial}$ in Eq.~\ref{eq:fbinomial} for the 250~pps pulse train is shown in black in Fig.~\ref{fig:Fano} and follows the simulation results.  At 5000~pps, there are strong effects of stimulus and spike history, so the binomial analogy is not expected to approximate the behavior of the model.  However, it does provide an upper bound for the possible Fano factors that the model can achieve, which is shown by the gray line.  This illustrates that there is room for the model to generate higher Fano factor values in response to high pulse rate stimulation, but different parameter values must be chosen.  To test the capacity for the model to produce higher Fano factors, we weakened the refractory effect by decreasing the time scale of the relative refractory period, $\tau_{\theta}$ in Eq.~\ref{eq:thresholdhistory}, to 200 $\mu s$.  This value is still within the range observed in physiological recordings \citep{Miller2001bi}.  All other model parameters were kept the same.  The results from simulations of this model, shown in green, illustrate that it can achieve higher Fano factor values, although the difference between Fano factors at the low and high pulse rates remains larger in the data.  

We were able to explore the relationship between refractory dynamics and Fano factor in the model more fully by developing a discrete-time Markov chain approximation to the point process model.  The Markov chain framework extends the Bernoulli process analogy (introduced above for low pulse rate stimulation) for stimuli in which spike history and interpulse interaction effects are present.  We used the Markov chain approximation to obtain estimates of firing rate and Fano factor and explore the full range of possible behaviors in the model.  See the Appendix for further details.

\subsection{Model predictions: amplitude-modulated pulse train stimuli}
\label{subsec:modelpredictam}

Modern cochlear implant speech processors provide temporal information to cochlear implant listeners by modulating the current levels of pulses over time.  It is important, therefore, to explore how the model responds to non-constant pulse trains.  In order to relate model results to available physiological and psychoacoustic data, we simulated responses to sinusoidally amplitude-modulated pulse trains.
Specifically, as inputs to the model we used trains of biphasic pulses with the current level of the $n^{th}$ pulses defined by the equation:
\begin{align}
I_n = \bar{I}(1+ m \sin(2 \pi t_n f_m)). \notag
\end{align}	
$I_n$ is the current level of the pulse with onset at time $t_n$, and the parameters $m$ and $f_m$ parameterize the modulation depth and modulation frequency, respectively.  As in previous sections, these pulses were charge balanced and had phase duration of 40~$\mu s$.   To compare the model with available physiological data \citep{Litvak2003sinusoid} and a recent modeling study \citep{OGorman2010}, we used a 5000~pps carrier pulse rate modulated at a frequency of 417 Hz.  We simulated two mean current levels, one that evoked approximately 50 spikes per second when the pulse train was unmodulated and a second that evoked approximately 100 spikes per second.  The duration of all pulse trains used in simulations was ten seconds, and the results from ten repeated simulations for each stimulus condition were used to compute standard errors in estimated response statistics.

\begin{figure}[t!]
\includegraphics[width=84mm]{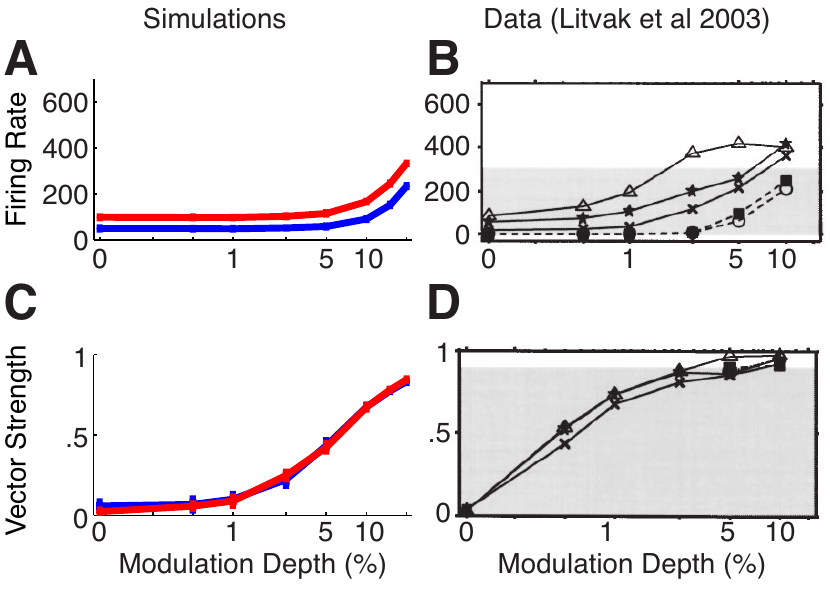} 
\caption{ Responses to sinusoidally amplitude-modulated pulse trains with 417 Hz modulation frequency.
\textbf{A,C:}  Simulation results obtained from responses to ten second long biphasic pulse trains with 40~$\mu s$ pulse durations and modulation frequency given on the x-axis.  Two mean current levels were used, one that produced a mean firing rate of 50 spikes per second for unmodulated input (blue) and a higher value that produced a mean firing rate of 100 spikes per second (red).  Error bars represent standard error in the mean of ten repeated simulations.  VS is computed with reference to the period of modulation.
\textbf{B,D:}  Firing rate and VS measures obtained from five cat AN fibers stimulated by intracochlear monopolar cochlear implant stimulation using biphasic pulse trains with duration of 25 $\mu s$ per phase.  Reprinted with permission from \cite{Litvak2003sinusoid}, copyright Acoustical Society of America.
}
\label{fig:AMRateVS}
\end{figure}

In Fig.~\ref{fig:AMRateVS}A, we show simulated firing rates, as a function of modulation depth.  Firing rates increase with modulation depth for $m$ greater than approximately 5$\%$.  We compared these results to cat single fiber data measured by Litvak and colleages \citep{Litvak2003sinusoid}, which we have reproduced in Fig.~\ref{fig:AMRateVS}C.  The simulation results qualitatively match the experimental data, especially when compared to the two fibers that have lower firing rates (circle and filled square in C).  The main discrepancy between the simulated and recorded firing rates is that the model does not exhibit increased spiking until the modulation depth increases beyond $5\%$, whereas the recorded firing rates increase at modulation depths as low as $1\%$. 

To quantify the sensitivity of spike timing to the period of the modulated waveform, Litvak et al computed VS as in Eq.~\ref{eq:VS}.  The relevant period in this case is the period of modulation, so $T$=2.4ms in Eq.~\ref{eq:VS}.  Simulation results using the point process model are shown in Fig.~\ref{fig:AMRateVS}C and the experimental measurements of  \cite{Litvak2003sinusoid} are reproduced in panel D.  The model again qualitatively captures the increase in VS with increasing modulation depth, although with important quantitative differences.  In particular, the experimental data show that AN fibers have exquisite sensitivity to weak modulations, producing VS values of approximately 0.5 for modulation depths of only 0.5$\%$.  As shown by \cite{OGorman2010}, this sensitivity to weak modulations may be a consequence of nonlinear mechanisms in spike generators that are accounted for in Fitzhugh-Nagumo dynamical models, but appear to be lacking in this point process description.

\subsection{Application to amplitude modulation detection}
\label{subsec:pctcorrect}

\begin{figure}[t!]  
\includegraphics[width=84mm]{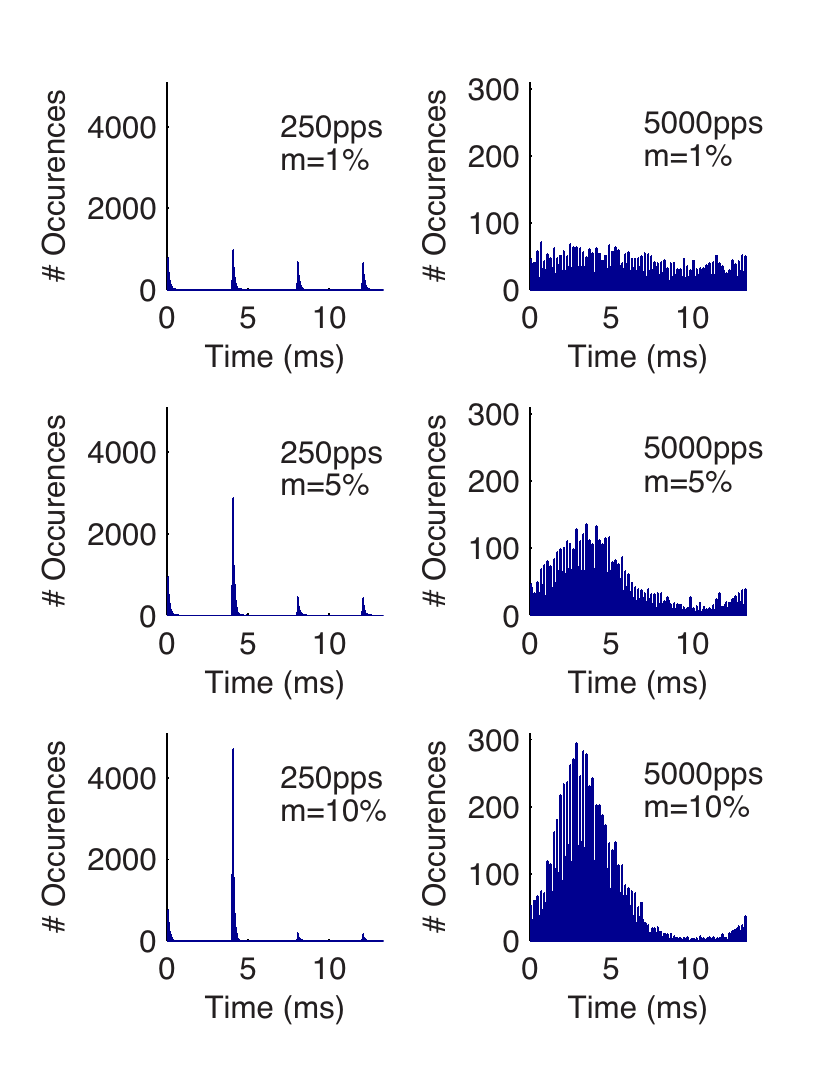} 
\caption{ Histograms of response to 250~pps (left column) and 5000~pps (right column) pulse trains modulated at 75 Hz.  Modulation depth increases from 1\% (top row) to 10\% (bottom row). Histograms were estimated from 10000 repeated simulations of responses to one cycle of trains of biphasic pulses, and plotted with $50~\mu s$ time bins.  Current levels were set separately for each pulse rate so that unmodulated pulse trains presented would evoke approximately 50 spikes per second.}
\label{fig:Periodogram}
\end{figure}

In this final set of simulations, we explore how the temporal information in spike patterns can be quantified in a way that enables simulation results to be interpreted in the context of psychoacoustic experiments studying the ability of cochlear implant listeners to detect modulation at varying carrier pulse rates \citep{Galvin2005, Pfingst2007, Galvin2009}.  We begin with a simple observation -- a 250 pps carrier, due to the tight phase-locking of spikes relative to the timing of pulses (see Fig.~\ref{fig:VS}A), does not appear to evoke AN activity that represents a modulated envelope.  This point is illustrated in the left column of Fig.~\ref{fig:Periodogram}.  Here, we show histograms of spike times obtained from one cycle of a pulse train modulated at 75 Hz.  From top to bottom, the modulation depth is increased from 1\% to 10\%.  The current level is fixed so that 50 spikes per second are evoked, on average, in response to unmodulated stimuli.  Our intuition tells us that this sparse and punctate pattern of spikes will not effectively transmit information about a slowly-varying envelope.  Nonetheless, these responses still carry some information about modulation depth, as evidenced, for instance, by the increasing height of the second peak as modulation depth increases.

In the right column of Fig.~\ref{fig:Periodogram}, we show responses to a cycle of a 5000~pps pulse train modulated at 75 Hz.  These histograms indicate that the high rate carrier appears to provide a more complete representation of the envelope of the modulated stimulus.  In contrast to the low rate carrier, then, we may expect that this carrier would provide greater temporal information regarding the presence of a sinusoidal modulation.  In other words, we may expect improved amplitude modulation detection if an observer (for instance higher processing centers in the auditory pathway) had access to this pattern of spikes as opposed to those in the left column.

\begin{figure}[t!]  
\includegraphics[width=84mm]{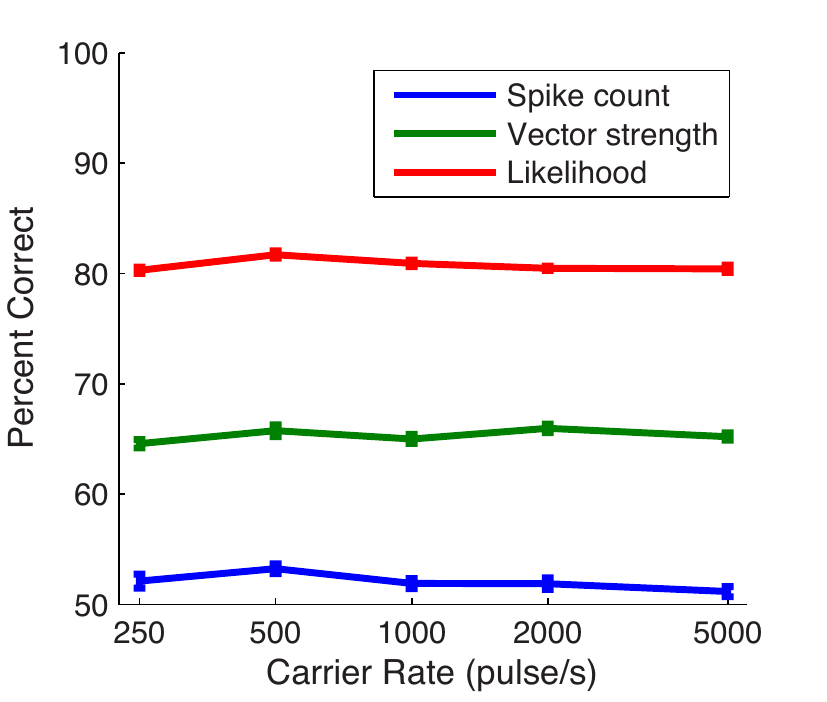} 
\caption{ Percent discrimination of a sinusoidally amplitude-modulated pulse train.  Three discrimination measures are were used to discriminate unmodulated pulse train from a modulated pulse train on the basis of simulated spike trains (see text for details).  Both stimuli were one second long trains of biphasic pulses, 40$\mu s$ per phase and current level was varied for each carrier pulse rate so that the unmodulated pulse train always evoked approximately 50 spikes per second, on average.  Modulated pulse trains had 75 Hz modulation frequency and 1\% modulation depth.  Percent correct values were computed from 1000 repeated simulations of the model, error bars represent standard errors in the mean of ten repeated estimates of percent correct. Chance level is 50\% correct.}
\label{fig:AMPctCorrect}
\end{figure}

To test these statements quantitatively, we used an ideal observer analysis to simulate modulation detection based on spike trains generated by the point process model.
A general result of point process theory allows us to use the conditional intensity function $\lambda(t | I(t), H(t))$ in Eq.~\ref{eq:cifmodel} to compute the log-likelihood of observing a sequence of spike times $\{ t_i \}_i^N$, conditioned on the input stimulus $I(t)$ \citep{Daley2003, Paninski2004pp, Truccolo2005}:
\begin{subequations}
\begin{align}
L(\{t_i\}_1^N | I) = \sum_{i=1}^N &\log(\lambda(t_i | I(t_i),H(t_i)) - \notag \\ 
& \int_0^s \lambda(t|I(t),H(t))dt, \tag{\ref{eq:likelihood}}
\end{align}
\label{eq:likelihood}
\end{subequations}
where $s$ is the total duration of the stimulus.
To analyze how temporal modulations are encoded in spike trains, we simulated two spike trains: one in response to an unmodulated pulse train and one in response to a modulated pulse train with 75 Hz modulation frequency and 1\% modulation depth.  Both stimuli were one second long and consisted of biphasic pulses that were 40$\mu s$ per phase.  After computing the likelihood function (Eq.~\ref{eq:likelihood}) for all possible pairings of spike trains and stimuli, we used a likelihood ratio test to discriminate between the two spike trains \citep{Green1966, Pillow2005, Goldwyn2010jcn}.  If the true pairings of stimulus and response produce the highest likelihood values, then the ideal ideal observer is said to correctly detect the modulated stimulus.  By repeating this procedure, we estimated percent correct detection values.  We also varied the carrier pulse rate from a low carrier pulse rate (250~pps) to a high carrier pulse rate (5000~pps) to investigate our previous observation that low pulse rate stimulation produces an (apparently) incomplete representation of the modulated waveform.

The results of this analysis are shown by the red line in Fig.~\ref{fig:AMPctCorrect}, where error bars indicate the standard error in the mean of 10 computed values of percent correct.  The x-axis shows the carrier pulse rate used as the input to the point process model.  At all pulse rates tested, for this controlled spike rate of 50 spikes per second, this maximum likelihood discrimination procedure produced correct discrimination on approximately 80\% of trials.  These simulations did not reveal any strong changes due to the carrier pulse rate.  Despite the visible differences in Fig.~\ref{fig:Periodogram}, the spike patterns in response to modulated stimuli are equally detectable regardless of the carrier pulse rate.

To further characterize the information in these spike patterns, we also computed discrimination measures using  decision rules that selected the modulated pulse train on the basis of which spike train response had a higher spike count (blue line) or a higher vector strength with reference to the 75Hz modulation (green line).  From Fig.~\ref{fig:AMRateVS}, we can see that spike count and vector strength increase with modulation depth, so simulated spike trains can encode the presence of modulation information using either of these cues.  For the small modulation depth used in these simulations, the spike count discrimination rule had near chance performance at just over 50\% correct.  The vector strength discrimination measure produced a higher percent correct, although it was still substantially below the value obtained using the spike train likelihood technique discussed above, indicating that vector strength as a measure of phase locking does not completely describe how temporal modulations are expressed in the precise sequence of spike times.   Similar to the likelihood function discrimination measure, neither spike count nor vector strength discrimination predicted percent correct values that depended strongly on the carrier pule rate.  

To summarize these simulations of modulation detection: we found that modulation detection based on three different decision rules did not vary across a range of carrier pulse rates.  This ideal observer analysis suggests that modulation detection does not require peripheral spike patterns that fully represent the sinusoidal envelope of these pulse trains.

\section{Discussion}
\label{sec:discussion}

\subsection{Review of main findings}
\label{subsec:reviewfindings}
In this study, we have used point process theory to derive mathematical expressions for statistical measures of neural excitability that are commonly used to characterize the response properties of AN fibers to electrical stimulation.   Furthermore, we proposed an explicit model with features that could be related to biophysical mechanisms in a phenomenological sense. These included temporal filtering that represented subthreshold dynamics of the membrane potential, a nonlinearity associated with spike generating processes, and a secondary filter that accounts for variability in spike timing.    

An essential feature of the proposed modeling framework is that all parameters can be determined from responses to single and pairs of electric pulses.  The model is minimal in the sense that each model parameter is uniquely identified with a single statistic reported in physiological experiments, as illustrated in Fig.~\ref{fig:param}.   Moreover, the point process model incorporated dynamical and stochastic properties that are relevant to high pulse rate stimulation.  Specifically, the relative spread of the model depends on the time since the last spike in a manner consistent with data reported in \citep{Miller2001}, and subthreshold pulses presented in rapid succession could combine to increase the excitability of the model neuron, a facilitation phenomenon observed in electrical stimulation of AN fibers \citep{Cartee20001, Cartee2006, Heffer2010}.  

The construction of the model ensured that it would reproduce the measures of threshold, relative spread, jitter, chronaxie, as well as changes in threshold and relative in the context of response to pairs of pulses separated by an interpulse interval.  To further test whether our model generalizes, we simulated responses to extended stimuli of greater relevance to cochlear implant stimulation: pulse trains with constant current per pulse and current levels that were sinusoidally amplitude-modulated.  The model produced interspike interval distributions that qualitatively agreed with data reported in \citep{Miller2008changes} (see Fig.~\ref{fig:RateAndISI}), and also provided important insight into how temporal jitter in spike times could account for synchronization to pulse trains presented at multiple pulse rates (see Fig.~\ref{fig:VS}).  Overall, it appears that the proposed point process framework provides a satisfactory description of AN spike trains, although there are important shortcomings and potential extensions to consider, which we discuss below.

\subsection{Relation to past modeling studies and future directions}
\label{subsec:relation}

The point process model represents an extension to previous simplified models of AN spiking.  In particular, the model developed in \cite{Bruce1999train, Bruce1999single} and related stochastic threshold crossing models \citep{Litvak2003sinusoid} by including temporal integration, facilitation, spike history dependent relative spread, and a realistic amount of spike time jitter.  The simplest dynamical models that describe AN responses to electric stimulation are integrate and fire models \citep{Stocks2002, Chen2007}.  As explained in Sec.~\ref{subsec:biophysical}, the point process model can be interpreted as a modified integrate and fire model with escape noise.  One significant difference between our approach and standard integrate and fire models is that we have introduced an asymmetry in the otherwise linear subthreshold dynamics, enabling the model to exhibit facilitation in response to charge balanced biphasic pulses.  An advantage of the point process model over standard integrate and fire models is that spiking probabilities can be computed relatively simply using the conditional intensity function and the corresponding lifetime distribution function in Eq.~\ref{eq:Lifetime}.  One does not need to estimate hazard rates \citep{Plesser2000} or solve difficult first passage time problems.  We have used this mathematical tractability to derive a parameterization method on the basis of physiological data.  

We suggest that this feature is an important strength of our model.  Cochlear implants stimulate thousands of AN fibers, and response statistics such as threshold, relative spread, and jitter can vary widely across the population of AN fibers, see for instance \citep{Miller1999, Miller2001bi}.  Constructing populations of AN fibers with a representative distribution of thresholds has been done using a biophysically detailed model \citep{Imennov2009}, but the point process framework provides an explicit method for controlling a number of key response properties of model neurons.  In addition to facilitating future studies that consider the response properties of populations of AN fibers, this parametric control can also be used to investigate which features may impact sensory perception in a significant way.  For instance, physiological data from rat AN fibers suggest that long-term deafness can increase absolute refractory periods and excitability thresholds \citep{Shepherd2004}. Such changes can be incorporated into the model via straightforward modifications of parameter values.

Although the model accurately predicted responses to extended pulse trains in several ways, it was unable to quantitatively predict the precise dependence of Fano factor on pulse rate and firing rate (Fig.~\ref{fig:Fano}) and the exquisite sensitivity of AN fibers to weak modulations depths as small as 0.5\% (Fig.~\ref{fig:VS}).  An explanation for these results can be found in \cite{OGorman2010}, which showed that the Fitzhugh-Nagumo exhibits a dynamical instability when driven with 5000~pps stimuli, and that this mechanism can account for experimentally measured values of Fano factor and strong phase locking to weak modulations reported by \cite{Litvak2003sinusoid}.  Future work could seek to synthesize these developments by developing models that contain this dynamical mechanism and can also be both parameterized to additional physiological data and directly used in likelihood-based discrimination studies.  One possible approach for improving the sensitivity of the model to small modulation is to apply an additional filter to the stimulus that acts as a differentiator on the sinusoidal evelope.  Filters could also be included, for instance, to account for resonator dynamics that may affect AN responses to cochlear implant stimulation \citep{Macherey2007}.  The analogy of subthreshold dynamics with escape noise discussed in Sec.~\ref{subsec:biophysical} can provide useful insight into the effects of additional filters on the point process model.  

An alternative approach, that would still be grounded in the point process framework, would be to fit a model directly to a richer set of stimuli that are more relevant to the study of cochlear implant speech processing strategies.  We have pointed out that the form of the model presented here, a cascade of linear and nonlinear stages, is similar to standard GLMs \citep{Paninski2004pp}.  GLMs have been shown to capture the activity of sensory neurons in retina with a high degree of temporal precision and can be fit from sets of spike trains recorded in response to arbitrary time-varying stimuli~\citep{Pillow2005}.  They have also been recently applied to auditory nerve recordings for acoustic stimulation \citep{Trevino2010, Plourde2011}. An important direction for future work, therefore, would be to fit GLMs to single unit recordings of AN fibers using stimulus sets that are clinically relevant to cochlear implant speech processing.  One could also pursue a model-based approach and fit GLMs to biophysically detailed models of AN responses to cochlear implant stimluation \citep{Imennov2009, Woo2010}.  The resulting point process descriptions would not necessarily be constrained by data from single pulse and paired pulse stimuli, but the mathematical relationships in Sec.~\ref{subsec:pptheory} could still be used to analyze the resulting models.  

Incorporating firing rate adaptation is another important direction for future models of AN responses to cochlear implant stimulation.  Recordings from AN fibers in deafened cats have shown that adaptation affects neural responses to amplitude-modulated cochlear implant stimuli \citep{Hu2010} and differentially affects responses to low and high rate pulse trains \citep{Zhang2007}. GLMs can exhibit firing rate adaptation through the addition of longer lasting spike history effects. Computational modeling can provide further insight into the origins and coding consequences of adaptation for stimulation strategies \citep{Woo2010}.

\subsection{Implications for high pulse rate stimulation strategies}
\label{subsec:implications}

A motivation for this work was to develop a model that could accurately predict responses to high carrier pulse rate stimulation.  We have done this by including dynamical and stochastic features that reflect how past stimulation and past spiking activity can influence the excitability of AN fibers.   Our simulation results in Fig.~\ref{fig:Periodogram}, past modeling studies \citep{Rubinstein1999, Mino2002, Chen2007}, and physiological evidence \citep{Litvak2003vowel, Litvak2003sinusoid} suggest that responses to high pulse rate stimulation may represent the envelope of temporally modulated stimuli with greater fidelity than responses evoked by low pulse rate stimulation.  Moreover, one would intuitively expect high pulse rate stimulation to provide greater temporal information to cochlear implant listeners and thereby improve speech perception and psychophysical performance on tests of temporal resolution.  The fact that, to date, no psychoacoustic studies have shown clear evidence that high pulse rate stimulation improves speech reception or amplitude modulation detection poses an intriguing challenge to our understanding of the connection between AN spiking activity and auditory perception.  

To connect AN spiking activity to psychoacoustic experiments, one can leverage the mathematical theory of point processes which provides access to tools from signal detection and information theory \citep{Heinz2001a, Heinz2001b, Goldwyn2010jcn}.  In Sec.~\ref{subsec:pctcorrect} we used the likelihood function of the point process model (Eq.~\ref{eq:likelihood}) to simulate modulation detection for a 75 Hz sinusoidally amplitude-modulated pulse train.  In human listeners, modulation detection at this frequency is correlated with speech perception \citep{Won2011}, so it is of interest to understand how neural responses transmit the temporal information in these stimuli.  

The simulated spike trains evoked by low pulse rate stimulation were equally discriminable to those evoked by high pulse rate stimulation in our simulation of amplitude modulation detection (Fig.~\ref{fig:AMPctCorrect}).  These results suggest the possibility that cochlear implant listeners could successfully discriminate modulated and unmodulated stimuli in the context of psychoacoustic experiments, even when patterns of evoked neural activity in the auditory periphery provide distorted or incomplete representation of the stimulus envelope.  This observation is consistent with psychoacoustic evidence that modulation detection in human listeners does not improve with high carrier pulse rate stimulation \citep{Galvin2005, Pfingst2007, Galvin2009}. Moreover, it highlights an essential challenge for improving listening outcomes for cochlear implant users -- in order to identify the relative benefits (or shortcomings) of high pulse rate stimulation, research should seek to identify psychoacoustic measures that reveal the perceptual consequences of the distinct patterns of neural activity evoked by low and high rate pulsatile stimulation.

\section*{Acknowledgments}
This work has been supported by NIDCD grants F31 DC010306 (JHG),  R01 DC007525 (JTR), and the Burroughs Wellcome Fund (ESB).

\section*{Appendix: Markov chain approximation to the point process model}
\label{sec:appendix}

\subsection*{Definition of the Markov chain for constant pulse train stimulation}
\label{subsec:appendix:definition}
In Sec.~\ref{subsubsec:irregularity}, we discussed how the firing rate and Fano factor produced by low pulse rate, constant current level stimulation can be estimated for the point process model by making an analogy to a Bernoulli process.  In this Appendix, we generalize this type of approximation for the case of high pulse rate, constant current level stimulation by using a Markov chain model to account for the effects of past spikes and stimulus history.  Results from Markov chain theory allow us to characterize the range of firing rates and firing irregularity (Fano factor) that the model can produce for this class of stimuli.

As in the Bernoulli process analogy, we simplify the problem by associating to each pulse a probability that the neuron spikes in response.  This probability is given by the lifetime distribution function in Eq.~\ref{eq:Lifetime}, where the integral is evaluated over a duration of one interpulse interval.  This probability value depends on the history of past pulses and past spikes. If we neglect precise spike timing, and presume that all spike times coincide with the onset of a pulse, then we can approximate the probabilities at hand via a sequence of values $\left\{  p_n(I) \right \}$, where the subscript $n$ represents the number of pulses that have elapsed since the last occurrence of a spike.
In the absence of history effects, for instance for low pulse rate stimulation, $p_n(I)$ is identical for all pulses since the interpulse interval is long relative to summation and refractory effects.  

\begin{figure}[!]
\includegraphics[width=75mm]{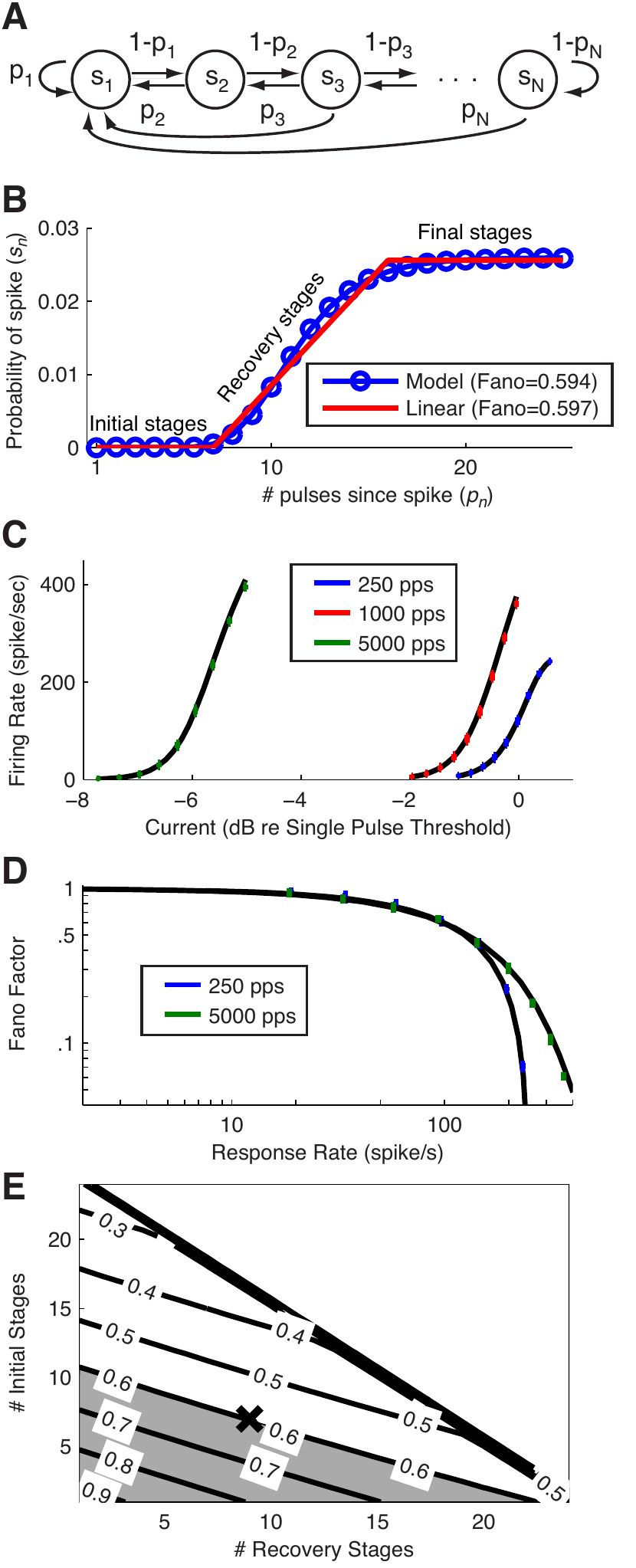} 
\caption{\footnotesize  Markov chain approximation to the point process model.
\textbf{A:} Schematic of the Markov chain, with $p_n$ representing the probability that the neuron spikes $n$ pulses after the previous spike.
\textbf{B:} Example of $p_n$ as a function of $s_n$ for 5000~pps stimulation set to evoke 100 spikes per second (blue).  Corresponding piecewise linear approximation (red), see text for details.
\textbf{C:} Firing rates of the point process model predicted by the Markov chain model.
\textbf{D:} Fano factor of the point process model predicted by the Markov chain model.
\textbf{E:} Dependence of the Fano factor of the piecewise linear approximation to the Markov chain model on the number of initial and recovery stages.  Contours show Fano factor values with firing rate set to be 100 spikes per second and pulse rate of 5000~pps.   Gray area indicates regions where Fano factor is higher at 5000~pps than the value obtained at 250~pps, and $\times$ marks the position of the point process model with parameter values given in Table~\ref{table:parameter}.
}
\label{fig:Appendix}
\end{figure}

To account for history effects, we define a discrete time Markov chain.  The states of this chain, which we denote $s_n$, represent the number of pulses that have elapsed since the previous spike.  There are two possible transitions away from each state.  If a spike occurs, then the chain returns to $s_1$.  The transition probability of this event is $p_n(I)$.  If no spike occurs, then the state of the chain advances by one to $s_{n+1}$.  The probability of this event is $1-p_n(I)$.  In practice, there is some number of pulses beyond which the refractory and summation effects no longer alter the probability of a spike, so we can limit the number of states in the chain to some finite number $N$.   A schematic illustration of the resulting Markov chain is shown in Fig.~\ref{fig:Appendix}A. An example of the sequence of transition probabilities $\left\{  p_n(I) \right \}$ is shown in Fig.~\ref{fig:Appendix}B for a 5000~pps stimulus that produces a firing rate of 100 spikes per second.  There are 25 total states in this chain (the x-axis), this indicates that history effects in the model do not persist beyond $\sim5$ms.  We therefore used a 5 state Markov chain for 1000~pps stimuli and a 2 state Markov chain for 250~pps stimulation.

\subsection*{Calculation of the firing rate and Fano factor}
\label{sub:appendix:calc}
Results from the theory of discrete-time, discrete-space Markov chains allow us to analytically compute the firing rate and Fano factor of the Markov chain approximation, for arbitrary pulse rates and current levels.  To do this, we use the fact that the Markov chain has a stationary distribution $\pi$ which gives the long time probability that the Markov chain will be in each state.  The stationary distribution can be obtained by computing the eigenvector associated with the (unique) eigenvalue that takes the value one \citep{Kemeny1960}.  In other words, the stationary distribution $\pi$ solves
\begin{align}
\pi M = \pi, \notag
\end{align}
where $M$ is the transition matrix for the Markov chain.  The first element of $\pi$, which we denote by $\pi_1$, represents the long-term proportion of time steps in which the chain is in state $s_1$.  Since the chain only returns to state $s_1$ if there has been a spike in response to the previous pulse, $\pi_1$ can be used to compute the firing rate. In particular, if we denote the pulse rate of the stimulus by $\rho$, then the firing rate (in spikes per second) is given by:
\begin{align}
\label{eq:apprate}
r =  \rho \pi_1 \;.
\end{align}
Fig.~\ref{fig:Appendix}C compares this Markov chain approximation for firing rate (black lines) to firing rates obtained from simulations of the point process model (colored error bars).  The simulated firing rates are the same as shown in Fig.~\ref{fig:RateAndISI}A.  The fact that the Markov chain approximation accurately predicts firing rates for the point process model at low pulse rates is expected because the spiking can be described by a Bernoulli process in this case. Importantly, however, the Markov chain also accurately approximates firing rates for the high pulse rate stimuli.  This framework, therefore, adequately captures the influence of spike history and interpulse effects in the point process model and can be used to  estimate the firing rate of the point process model via Eq.~\ref{eq:apprate}.

Next, we show how the Markov chain approximation to the point process model can be used to compute the Fano factor of the point process model for pulse train stimuli with constant current strength.  If we let $N(T)$ be the number of spikes produced over a period of time of length $T$, then the Fano factor is defined as:
\begin{align}
F =\frac{ \mathrm{Var}[N(T)] }{\mathrm{E}[N(T)]}. \notag
\end{align}
The denominator is given via the stationary distribution, as discussed above.  In particular, $\mathrm{E}[N(T)] = \pi_1 n_p$, where $n_p$ is the total number of pulses in the stimulus. To estimate the variance of $N(T)$, we first define the characteristic matrix \citep{Kemeny1960}
\begin{align}
 Z = \left[ Id - M+M_\infty \right ]^{-1} \notag
\end{align}
where $Id$ is the identity matrix, $M$ is the transition matrix for the Markov chain, and $M_\infty$ is a matrix whose rows are the vector $\pi$.  We then appeal to the law of large numbers for Markov chains, which states that the limiting variance of $N(T)/\sqrt{n_p}$ is $\pi_1 (2Z_{11} - \pi_1 - 1)$ \citep{Kemeny1960},
where $Z_{11}$ is the first entry of the characteristic matrix. 

To compute the Fano factor, we then take the ratio of the variance of $N(T)/\sqrt{n_p}$ to its expected value and find that the Fano factor is:
\begin{align}
\label{eq:FMC}
F =2Z_{11} - \pi_1 - 1. 
\end{align}
Fig.~\ref{fig:Appendix}D shows values of the Fano factor for 250~pps and 5000~pps stimulation computed using the equation for a range of firing rates (black lines).  Fano factor values computed for the point process model are shown with colored error bars, and are the same as those in Fig.~\ref{fig:RateAndISI}B.  Overall, there is close agreement between the Markov chain approximation given by Eq.~\ref{eq:FMC} and the Fano factor values computed from simulations of the point process model.  In sum, the Markov chain approximation accurately captures both the mean firing rate and the normalized variance of the spike count for a range of stimulus levels and pulse rates.

\subsection*{Piecewise-linear approximation to the Markov chain}
\label{sub:appendix:linear}

As discussed in the text, physiological data suggest that Fano factors measured from responses to 5000~pps pulse trains may be higher than those measured from responses to 250~pps pulse trains \citep{Miller2008changes}.  We were interested in identifying whether the point process model could exhibit the same behavior, so we used the Markov chain to systematically explore the space of possible Fano factors that the model could produce.  To do this, we first observed that for the 5000~pps pulse train, the $p_n(I)$ curves can be roughly caricatured as having three stages: an \emph{initial stage} where the probability of spiking is near zero, a \emph{recovery stage} where $p_n(I)$ increases in a relatively linear manner with $n$, and a \emph{final stage} where the $p_n(I)$ is nearly constant with $n$.  This piecewise linear approximation is shown in red in Fig.~\ref{fig:Appendix}B.  We therefore characterized the functions $p_n(I)$ by two parameters, one that describes the number of initial stages and one that describes the number of recovery stages.  We then swept across this space of two parameters to explore the range of possible Fano factor values for a fixed firing rate.  

An example of this procedure is shown in Fig.~\ref{fig:Appendix}E, where firing rate is set to be 100 spikes per second and the pulse rate is 5000~pps.  We show Fano factor values computed using the Markov chain model with the piecewise linear $p_n(I)$, and indicate in gray the regions of parameter space in which Fano factors are higher when stimulating the neuron with 5000~pps than with 250~pps pulse trains.  The black $\times$ indicates the location of the model with parameter values given in Table~\ref{table:parameter}.  In order to generate Fano factors that are larger for responses to 5000~pps stimuli than 250~pps stimuli, either the number of initial stages or the number of recovery stages must be decreased.  One mechanism in the model that controls the number of initial and recovery stages is the time scale of the threshold refractory effect, $\tau_\theta$ in Eq.~\ref{eq:thresholdhistory}.  Shortening the refractory period reduces the number of initial and recovery stages, and thus our analysis of the Markov chain model provides a theoretical basis for the observation in Fig.~\ref{fig:Fano} that the point process model with a smaller $\tau_\theta$ will have higher Fano factor for 5000~pps stimulation than 250~pps stimulation. The number of initial and recovery stages is also influenced by the time scale of stimulus integration, $\tau_\kappa$ in the stimulus filters $K^+$ and $K^-$ in Eqs.~\ref{eq:model2} and~\ref{eq:model3}.  The Markov chain approximation is therefore a useful tool for connecting between spike train statistics of the point process model to the biophysical properties (refractory effects, membrane time constant, e.g) that are represented by parameters in the model.

\newpage

\bibliographystyle{plainnat}     

\end{document}